\documentclass[%
 aps,
 prd,%
 amsmath,amssymb,
 reprint,showpacs,
nofootinbib
]{revtex4-1}

\usepackage{graphicx}% Include figure files
\usepackage{dcolumn}% Align table columns on decimal point
\usepackage{bm}% bold math
\usepackage{epstopdf}
\usepackage{color}
\usepackage[normalem]{ulem}
\usepackage{natbib} 

\usepackage{epstopdf}
\usepackage{subfigure}
\usepackage{wrapfig}

\usepackage{ifthen}
\usepackage{tikz}
\usepackage{tikz-3dplot}
\usetikzlibrary{intersections}
\begin{document}

%\preprint{AIP/123-QED}

\title[Spherically-symmetric solutions in GR]{
Spherically-symmetric solutions in general relativity}
\author{Do Young Kim}
\email{dyk25@mrao.cam.ac.uk}
 \affiliation{Astrophysics Group, Cavendish Laboratory, JJ Thomson Avenue, Cambridge CB3 0HE}%Lines break automatically or can be forced with \\
 \affiliation{Kavli Institute for Cosmology, Madingley Road, Cambridge CB3 0HA}
\author{Anthony N. Lasenby}%
 \email{a.n.lasenby@mrao.cam.ac.uk}
\affiliation{Astrophysics Group, Cavendish Laboratory, JJ Thomson Avenue, Cambridge CB3 0HE}%
\affiliation{Kavli Institute for Cosmology, Madingley Road, Cambridge CB3 0HA}
\author{Michael P. Hobson}
\email{mph@mrao.cam.ac.uk}
\affiliation{Astrophysics Group, Cavendish Laboratory, JJ Thomson Avenue, Cambridge CB3 0HE}%

%\date{\today}% It is always \today, today,
             %  but any date may be explicitly specified

\begin{abstract}
We present a tetrad-based method for solving the Einstein field equations for spherically-symmetric systems and compare it with the widely-used Lema\^{i}tre--Tolman--Bondi (LTB) model. In particular, we focus on the issues of gauge ambiguity and the use of comoving versus `physical' coordinate systems. We also clarify the correspondences between the two approaches, and illustrate their differences by applying them to the classic examples of the Schwarzschild and Friedmann--Robertson--Walker spacetimes. We demonstrate that the tetrad-based method does not suffer from the gauge freedoms inherent to the LTB model, naturally accommodates non-zero pressure and has a more transparent physical interpretation. We further apply our tetrad-based method to a generalised form of `Swiss cheese' model, which consists of an interior spherical region surrounded by a spherical shell of vacuum that is embedded in an exterior background universe. In general, we allow the fluid in the interior and exterior regions to support pressure, and do not demand that the interior region be compensated. We pay particular attention to the form of the solution in the intervening vacuum region and verify the validity of Birkhoff's theorem at both the metric and tetrad level.  We then reconsider critically the original theoretical arguments underlying the so-called $R_{\rm h} = ct$ cosmological model, which has recently received considerable attention. These considerations in turn illustrate the interesting behaviour of a number of `horizons' in general cosmological models.
\end{abstract}

% PACS, the Physics and Astronomy
                             % Classification Scheme.
%Use showkeys class option if keyword
                              %display desired
\maketitle

%\begin{quotation}
%The ``lead paragraph'' is encapsulated with the \LaTeX\
%\verb+quotation+ environment and is formatted as a single paragraph before the first section heading.
%(The \verb+quotation+ environment reverts to its usual meaning after the first sectioning command.)
%Note that numbered references are allowed in the lead paragraph.
%
%The lead paragraph will only be found in an article being prepared for the journal \textit{Chaos}.
%\end{quotation}

\section{Introduction\label{sec:intro}}

Spherically-symmetric solutions in general relativity are of fundamental importance to the study of compact objects, black holes and cosmology. Indeed, two of the oldest and most commonly studied exact solutions of Einstein's field equations are spherically symmetric: the Schwarzschild metric \cite{Schwarzschild1916} describes the gravitational field outside a static spherical massive body, and the Friedmann--Robertson--Walker (FRW) metric \cite{Friedmann1922,Friedmann1924,Lemaitre1931,Robertson1935,Robertson1936,Robertson1936a,Walker1937} describes a homogeneous and isotropic universe in terms of the evolution of its scale factor with cosmic time. Moreover, it was not long before McVittie \cite{McVittie1933,McVittie1956} combined the Schwarzschild and FRW metrics to produce a new spherically-symmetric solution that describes a point mass embedded in an expanding universe, although there still remains some debate regarding its physical interpretation \cite{Kaloper2010,Lake2011}.

Subsequently, there have been numerous studies of the general-relativistic dynamics of self-gravitating spherical systems. For example, Misner, Thorne \& Wheeler \cite{Misner1973} describe the spherically-symmetric collapse of a `ball of dust' having uniform density and zero pressure that is embedded in a static vacuum exterior spacetime, and later generalise their results to incorporate pressure internal to the object. By contrast, `Swiss cheese' models \cite{Harwit1998} consider an exterior expanding FRW universe, albeit pressureless, in which a uniform pressureless spherical object is embedded and surrounded by a `compensating void' that itself expands into the background and ensures that there is no net gravitational effect on the exterior universe.

A more realistic description than the Swiss cheese models is provided by models based on the Lema\^{i}tre--Tolman--Bondi (LTB) solution \cite{Lemaitre1933, Tolman1934, Bondi1947}. Such models can incorporate an arbitrary (usually continuous) density profile for the central object, which is usually not compensated but can be made so by an appropriate choice of initial radial density and velocity profiles. Nonetheless, these models again assume both the interior and exterior regions to be pressureless, although the LTB solution has recently been extended in \cite{Lasky2006} to describe a central object with pressure embedded in a static vacuum exterior, and in \cite{Lynden-Bell2016} to accommodate cosmological models with uniform pressure.

A recent resurgence of interest in Swiss cheese and LTB models has been prompted by the possibility that they may provide an explanation for observations of the acceleration of the universal expansion, without invoking dark energy. This might occur if we, as observers, reside in a part of the universe that happens to be expanding faster than the region exterior to it.  By observing a source in the exterior region, one would then measure an {\it{apparent}}\/ acceleration of the universe's expansion, but this would be only a local effect.  The effects of local inhomogeneities on the apparent acceleration of the universe have been widely studied \cite{Celerier2012, Celerier2012a, Bolejko2010, Bene2010, Kainulainen2009}, and have been linked with the observations of distant Type-Ia supernova. In addition, LTB models have been used to study the effects of inhomogeneities on observed cosmological parameters, such as the Hubble constant \citep{Romano2015,Romano2011,Romano2011a}, and to calculate effects of a void as a possible explanation for the cold spot in the cosmic microwave background (CMB) \citep{Szapudi2014,Nadathur2014}. 

The LTB model does, however, have some limitations. In addition to the usual restriction to pressureless systems, the LTB model is typically expressed in comoving coordinates and thus provides a Lagrangian picture of the fluid evolution that can be difficult to interpret. More importantly, the LTB metric contains a residual gauge freedom that necessitates the imposition of arbitrary initial conditions to determine the system evolution. 

As a consequence, we have for some time adopted a different, tetrad-based method for solving the Einstein field equations for spherically-symmetric systems.  The method was originally presented in \cite{Lasenby1998} in the language of geometric algebra, and was recently re-expressed in more traditional tetrad notation in \cite{NLH1,NLH3}. The advantages of the approach are that it can straightforwardly accommodate pressure, has no gauge ambiguities (except in vacuum regions, as we shall discuss later) and is expressed in terms of a `physical' (non-comoving) radial coordinate. As a result, in contrast to the LTB model, the method has a clear and intuitive physical interpretation. Indeed, the gauge choices employed result in equations that are essentially Newtonian in form.

In \cite{Lasenby1999, Dabrowski1999}, we applied the method to modelling the evolution of a finite-size, spherically-symmetric object with continuous radial density and velocity profiles that is embedded in an expanding background universe (either spatially-flat, open or closed) and compensated so that it does not exert any gravitational influence on the exterior universe; the fluid was assumed to be pressureless throughout. In \cite{NLH1}, we used the method to obtain solutions describing a point mass residing in either a spatially-flat, open or closed expanding universe containing a cosmological fluid with pressure. In the spatially-flat case, a simple coordinate transformation relates our solution to the corresponding one derived by McVittie, but for spatially-curved cosmologies our metrics differ from the corresponding McVittie metrics, and we believe the latter to be incorrect. In \cite{NLH3}, we extended this study by applying the tetrad-based approach to obtain the solution describing the evolution of a finite spherical region of uniform interior density that is embedded in a background of uniform exterior density, where the fluid in both regions can support pressure and the expansion (or contraction) rates of the two regions are expressed in terms of interior and exterior Hubble parameters that are, in general, independent. We also derived a generalised form of the Oppenheimer--Volkov equation, valid for general time-dependent, spherically-symmetric systems.

In this paper, we present a comparison of our tetrad-based methodology with the LTB model for solving the Einstein field equations for spherically-symmetric systems. In particular, we focus on the issues of gauge ambiguity and the use of comoving versus `physical' coordinate systems. We also clarify the correspondences, where they exist, between the two approaches. In addition, we extend the analysis presented in \cite{NLH3} by applying our tetrad-based method to a generalised form of `Swiss cheese' model, which consists of an interior spherical region surrounded by a spherical shell of vacuum that is embedded in an exterior background universe. In general, we allow the fluid in the interior and exterior regions to support pressure, and we demand neither that the interior region be compensated, nor that the interior and exterior regions be uniform. Nonetheless, our principal focus is the case in which the fluid in the interior and exterior regions has uniform (although, in general, different) densities. In particular, we pay special attention to the form of the solution in the intervening vacuum region and verify the validity of Birkhoff's theorem, the usual interpretation of which has recently been brought in question \cite{Zhang2012}.  This investigation allows us to reconsider critically the original theoretical arguments underlying the so-called $R_{\rm h} = ct$ cosmological model \cite{MeliaShevchuk2012}, which has recently received considerable attention. These considerations in turn elucidate the behaviour of a number of `horizons' during the general-relativistic evolution of a spherically-symmetric self-gravitating matter distribution, which does not appear to have been widely discussed in the literature.

The structure of this paper is as follows. In Section~\ref{sec:framework}, we outline our tetrad-based approach to solving the Einstein equations. In Section~\ref{sec:ltb} we compare our approach to the more commonly-used LTB model. We apply our tetrad-based method to describe the evolution of a generalised form of `Swiss cheese' model in Section~\ref{sec:gscm} and investigate the validity of Birkhoff's theorem in its vacuum region in Section~\ref{sec:birkhoff}. We discuss the $R_{\rm h} = ct$ cosmology in Section~\ref{sec:rctmodel} and describe the generic evolution of a number of cosmological `horizons' in Section~\ref{sec:horizons}. Finally, we present our conclusions in Section~\ref{sec:conc}. We adopt natural units $c=G=1$ throughout.  

\section{Tetrad-based solution for spherical systems}
\label{sec:framework}

%We now summarise our tetrad-based method for solving the Einstein
%field equations for spherically-symmetric systems,
%which was
%originally presented in \cite{Lasenby1998} in the language of
%geometric algebra and re-expressed in more traditional tetrad notation
%in \cite{NLH1,NLH3}.
In a Riemannian spacetime in which events are labelled with a set of coordinates $x^{\mu}$, each point has the corresponding coordinate basis vectors $\mathbf{e}_{\mu}$, related to the metric via $\mathbf{e}_{\mu}\cdot\mathbf{e}_{\nu}=g_{\mu\nu}$.  At each point we may also define a {\itshape{local Lorentz frame}} by another set of orthogonal basis vectors $\hat{\mathbf{e}}_a$ (Roman indices), which are not derived from any coordinate system and are related to the Minkowski metric $\eta_{ab}=\mbox{diag}(1,-1,-1,-1)$ via $\hat{\mathbf{e}}_a\cdot \hat{\mathbf{e}}_b=\eta_{ab}$.  One can describe a vector $\mathbf{v}$ at any point in terms of its components in either basis: for example $v_{\mu}=\mathbf{v}\cdot \mathbf{e}_{\mu}$ and $\hat{v}_a=\mathbf{v}\cdot\hat{\mathbf{e}}_{a}$.  The relationship between the two sets of basis vectors is defined in terms of tetrads, or {{vierbeins}} ${e_{a}}^{\mu}$, where the inverse is denoted ${e^{a}}_{\mu}$:
\begin{equation}
\hat{\mathbf{e}}_a={e_{a}}^{\mu}\mathbf{e}_{\mu},\qquad
\mathbf{e}_{\mu}={e^{a}}_{\mu}\hat{\mathbf{e}}_a.
\label{eqn:tetraddef}\\
\end{equation}
It is not difficult to show that the metric elements are given in terms of the tetrads by $g_{\mu\nu}=\eta_{ab}e^a_{\ \mu}e^b_{\ \nu}$.

The local Lorentz frames at each point define a family of ideal observers whose worldlines are the integral curves of the timelike unit vector field $\hat{\mathbf{e}}_0$. Along a given worldline, the three spacelike unit vector fields $\hat{\mathbf{e}}_i$ $(i=1,2,3)$ specify the spatial triad carried by the corresponding observer. The triad may be thought of as defining the orthogonal spatial coordinate axes of a local laboratory frame that is valid very near the observer's worldline. In general, the worldlines need not be time-like geodesics, and hence the observers may be accelerating.

The Einstein--Hilbert action for general relativity is invariant under general coordinate transformations and local rotations of the Lorentz frames, which together constitute the gauge freedoms at our disposal.  For a spherically-symmetric system, we start by introducing a set of spherical polar coordinates $[x^\mu] = (t,r,\theta,\phi)$ and their corresponding coordinate basis vectors $\mathbf{e}_\mu$ $(\mu=0,1,2,3)$.  We first demand that (minus) the angular part of the line-element $ds^2=g_{\mu\nu}\,dx^\mu\,dx^\nu$ has the form $r^2\,d\Omega^2$, where $d\Omega^2 = d\theta^2+\sin^2\theta\,d\phi^2$. Aside from trivial spatial rotations of the coordinates, which leave the description of the spherically-symmetric system unchanged, this choice absorbs the gauge freedoms associated with transformations of the $r$, $\theta$ and $\phi$ coordinates, and in particular lifts $r$ from the status of an arbitrary radial coordinate to a quantity that is, in principle, physically measurable. It is a `physical' (non-comoving) coordinate for which the proper area of a sphere of radius $r$ is $4\pi r^2$.

The next step is to determine the general form of the tetrad ${e_{a}}^{\mu}$ that is consistent with spherical symmetry and this choice of coordinates. One immediately requires that, in \eqref{eqn:tetraddef}, the coordinate basis vector pairs $\{\mathbf{e}_0, \mathbf{e}_1\}$ and $\{\mathbf{e}_2, \mathbf{e}_3\}$ decouple. Moreover, one can perform local rotations of the Lorentz frames to align $\hat{\mathbf{e}}_2$ and $\hat{\mathbf{e}}_3$ with the coordinate basis vectors $\mathbf{e}_2$ and $\mathbf{e}_3$ at each point. Consequently, the tetrad components ${e_{a}}^{\mu}$ may be written in terms of four unknown functions, which we denote by $f_1(r,t)$, $f_2(r,t)$, $g_1(r,t)$ and $g_2(r,t)$. Note that dependencies on both $r$ and $t$ will often be suppressed in the equations presented below, whereas we will usually make explicit dependency on either $r$ and $t$ alone.  In particular, we may take the non-zero tetrad components and their inverses to be
\begin{align}
{e_{0}}^{0}&=f_1,&
{e^0}_{0}&=g_1/(f_1g_1-f_2g_2),\nonumber\\
{e_1}^{0}&=f_2,&
{e^0}_1 &=-f_2/(f_1g_1-f_2g_2),\nonumber\\
{e_0}^{1}&=g_2,&
{e^1}_0&=-g_2/(f_1g_1-f_2g_2),\nonumber\\
{e_1}^{1}&=g_1,&
{e^1}_1 &=f_1/(f_1g_1-f_2g_2),\nonumber\\
{e_2}^{2}&=1/r,&
{e^2}_2 &=r,\nonumber\\
{e_3}^{3}&=1/(r\sin\theta),&
{e^3}_3&=r\sin\theta.\label{eq:tetrads}
\end{align}

Our remaining gauge freedoms lie in the ability to transform to a new time coordinate, which may be a function of $t$ and $r$, and in performing local rotations of the Lorentz frames in the $(\hat{\mathbf{e}}_0,\hat{\mathbf{e}}_1)$-hyperplane (corresponding to a Lorentz boost in the radial direction at each point). The former possibility gives us complete freedom in the choice of the function $f_2$, and the greatest simplification of the tetrad components \eqref{eq:tetrads} is obtained by setting $f_2 \equiv 0$, which we call the Newtonian gauge because it allows simple Newtonian interpretations of the dynamics, as we shall see. In this gauge, the metric coefficients derived from the tetrad components lead to the line-element
\begin{equation}
ds^2=\frac{g_1^{2}-g_2^{2}}{f_1^{2}g_1^{2}}\,dt^2+\frac{2g_2}{f_1g_1^{2}}\,dt\,dr-\frac{1}{g_1^{2}}\,dr^2-r^2d\Omega^2.\label{eq:generalmetric}
\end{equation}

Finally, the remaining gauge freedom (which leaves the line-element unchanged) can be employed (at least in non-vacuum regions) to choose the timelike unit frame vector $\hat{\mathbf{e}}_0$ at each point to coincide with the four-velocity of the fluid at that point. Thus, by construction, the four-velocity $\mathbf{v}$ of a fluid particle (or an observer comoving with the fluid) has components $[\hat{v}^a] = [1,0,0,0]$ in the tetrad frame. Since $v^{\mu}=e_a^{\ \mu}\hat{v}^a$, the four-velocity may be written in terms of the tetrad components and the coordinate basis vectors as $\mathbf{v}= f_1 \mathbf{e}_0 + g_2 \mathbf{e}_1$. Thus, the components of a comoving observer's four-velocity in the coordinate basis are simply $[v^\mu] \equiv [\dot{t},\dot{r},\dot{\theta},\dot{\phi}]=[f_1,g_2,0,0]$, where dots denote differentiation with respect to the observer's proper time $\tau$. 

As a consequence of this final gauge choice, it is convenient to define the two linear differential operators
\begin{align}
L_t&\equiv f_1\partial_t+g_2\partial_r,\nonumber\\
L_r&\equiv g_1\partial_r.\label{eq:operators}
\end{align}
We may identify $L_t$ as the derivative with respect to the proper time of a comoving observer, since $L_t = \dot{t}\partial_t + \dot{r}\partial_r = d/d\tau$, and similarly one may show that $L_r$ coincides with the derivative with respect to the radial proper distance of a comoving observer.  Moreover, since $g_2$ is the rate of change of the $r$ coordinate of a fluid particle with respect to its proper time, it can be physically interpreted as the fluid velocity. We will therefore, in general, use $g_2$ and $v$ interchangeably in our analysis.

It is also convenient to introduce explicitly the spin-connection coefficients $F \equiv {\omega^0}_{11}$ and $G \equiv {\omega^1}_{00}$, as described in \cite{NLH1}, which are both, in general, functions of $t$ and $r$. Since we are assuming standard general relativity, however, for which torsion vanishes, the spin-connection coefficients can be written entirely in terms of the tetrad components and their derivatives. For the torsion to vanish and for the resulting Riemann tensor to satisfy its Bianchi identity, the spin-connection coefficients $F$ and $G$ and the non-zero tetrad components $f_1$, $g_1$ and $g_2$ must satisfy the relationships
\begin{align}
L_rf_1&=-Gf_1\Rightarrow
f_1=\exp\left\{-{\textstyle\int^r\frac{G}{g_1}dr}\right\},\nonumber\\
L_r g_2 &= Fg_1 
%\hspace{0.28cm}\Rightarrow F = \partial_r g_2
,\nonumber\\
L_t g_1 &= Gg_2,\phantom{f_1=\exp\left\{-{\textstyle\int^r\frac{G}{g_1}dr}\right\}}
\label{eq:tetrels}
\end{align}
where the explicit solution for $f_1$ contains no arbitrary function of $t$, because one can always be absorbed by a further $t$-dependent rescaling of the time coordinate (which does not change $f_2$).

For matter in the form of a perfect fluid with proper density $\rho$ and isotropic rest-frame pressure $p$, the Einstein field equations and the contracted Bianchi identities lead to the following system\footnote{In \cite{Lasenby1998}, two further equations are given, namely $L_r g_1=F g_2+\frac{M}{r^2}-\tfrac{1}{3}\Lambda r-4\pi r\rho$ and $L_t g_2=G g_1-\frac{M}{r^2}+\tfrac{1}{3}\Lambda r-4\pi r p$, but these may be derived from the $L_rM$ and $L_tM$ equations, respectively, in combination with the definition of $M$ given in \eqref{eq:quantities}.} of dynamical and continuity equations
\cite{Lasenby1998}
\begin{align}
%L_rf_1&=-Gf_1\Rightarrow f_1=\exp\left\{-\int^r\frac{G}{g_1}dr\right\},\nonumber\\
%L_r g_1&=F g_2+\frac{M}{r^2}-\tfrac{1}{3}\Lambda r-4\pi r\rho,\nonumber\\
L_r p&=-G(\rho+p),\nonumber\\
L_rM&=4\pi g_1 r^2\rho,\nonumber\\
%L_t g_2&=G g_1-\frac{M}{r^2}+\tfrac{1}{3}\Lambda r-4\pi r p,\nonumber\\
L_t\rho&=-\left(\frac{2g_2}{r}+F\right)(\rho+p),\nonumber\\
L_t M&=-4\pi g_2 r^2p,\label{eq:alleqns}
\end{align}
where we have defined the function of $t$ and $r$ (in general)
\begin{equation}
M\equiv \tfrac{1}{2}r\left(g_2^{2}-g_1^{2}+1-\tfrac{1}{3}\Lambda
r^2\right),\label{eq:quantities}
\end{equation}
and $\Lambda$ is the cosmological constant. 

The physical interpretation of the functions $F$, $G$ and $M$ is straightforward. As shown in \cite{NLH1}, for an object in general radial motion (not necessarily co-moving with the fluid) with four-velocity components $[\hat{u}^a] = [\hat{u}^0,\hat{u}^1,0,0]$ in the tetrad frame, the corresponding components of the object's four-acceleration are
\begin{align}
\hat{a}^0 &=\dot{\hat{u}}^0+G\hat{u}^0\hat{u}^1+F(\hat{u}^1)^2,\nonumber\\
\hat{a}^1&=\dot{\hat{u}}^1+G(\hat{u}^0)^2+F\hat{u}^0\hat{u}^1,\label{eq:v2}
\end{align}
and its proper acceleration is $\alpha = \sqrt{-\hat{a}^b\hat{a}_b}$, which provides a physical interpretation of the functions $F$ and $G$.
%\footnote{As further discussed in \cite{NLH1}, for an idealised test
%  object that is assumed to be infinitesimal, and so not subject to
%  fluid forces due to pressure gradients, but only to gravitational
%  forces, the proper acceleration is equal to the force per unit mass
%  (provided, for example, by a rocket engine) required to keep the
%  particle in this state of motion.}
In particular, for the special case in which the object is co-moving with the fluid, one has $[\hat{u}^b] = [1,0,0,0]$ and so $[\hat{a}^b] = [0,G,0,0]$. Thus the proper acceleration of a fluid particle is $\alpha = G$ in the radial direction. Indeed, the $L_rp$-equation in \eqref{eq:alleqns} shows that, in the absence of a pressure gradient, $G$ vanishes and so the motion becomes geodesic. The physical interpretation of the function $M$ can be obtained from the forms of the equations in \eqref{eq:alleqns} in which it appears.  In particular, the $L_rM$-equation can be written simply as $\partial_rM=4\pi r^2\rho$, which shows that $M$ plays the role of an intrinsic mass that is determined by the amount of mass-energy in a sphere of radius $r$.

The equations \eqref{eq:tetrels}--\eqref{eq:quantities} thus have clear physical interpretations and contain no residual gauge freedom (in non-vacuum regions). In particular, given an equation of state $p=p(\rho)$, and initial data in the form of the density $\rho(r,t_0)$ and the velocity $g_2(r,t_0)$, the future evolution of the system is fully determined. This is because $\rho$ determines $p$ and $M$ on a time slice and the definition of $M$ then determines $g_1$. The equations for $L_rg_2$, $L_rp$ and $L_rf_1$ then determine the remaining information, namely $F$, $G$ and $f_1$ respectively, on the time slice. Finally, the $L_t\rho$ equation and $L_tM$ equation (together with the definition of $M$) enable one to propagate $\rho$ and $g_2$, respectively, to the next time slice and the repeat the process. The equations can thus be implemented numerically as a simple set of first-order update equations. This approach was illustrated in \cite{Lasenby1998,Lasenby1999,Dabrowski1999}.

An alternative way of solving the system of equations \eqref{eq:tetrels}--\eqref{eq:quantities}, which was employed in \cite{NLH1,NLH2,NLH3}, is not to impose an equation of state, but instead specify a form for $\rho(r,t)$ for all $t$ or, equivalently, a form for $M(r,t)$ followed by use of the $L_rM$. In general, the remaining equations need to be solved as a set of coupled PDEs. Nonetheless, as shown in \cite{NLH1,NLH2,NLH3}, if $\rho(r,t)$ is piecewise uniform in $r$, then one may combine the $L_t\rho$, $L_tM$ and $L_rM$ equations to obtain an ODE in $r$ that may be solved to obtain an expression for the fluid velocity $g_2(r,t)$ and hence $F(r,t)$, albeit with each containing a time-dependent `constant' of integration, and the definition of $M$ then determines $g_1(r,t)$. One may then obtain the fluid pressure $p(r,t)$ by first using the $L_tM$ equation to eliminate $f_1$ from the $L_rp$ equation, which then yields the `generalised Oppenheimer--Volkov' equation \cite{NLH3}
\begin{widetext}
\begin{equation}
\partial_r p =
 - \left(\frac{\rho+p}{r}\right)\frac{M+4\pi r^3p-\frac{1}{3}\Lambda r^3 + r^2 v\partial_r v - 4\pi r^4 (\rho+p)  (\partial_t M)^{-1}v\partial_t v}
{(1+v^2)r-2M-\frac{1}{3}\Lambda r^3}.\label{eq:OVgeneralised}
\end{equation}
\end{widetext}
This equation is, in fact, valid for any spherically-symmetric perfect fluid system and reduces to the standard Oppenheimer--Volkov equation with a cosmological constant \cite{Oppenheimer1939,winter2000} for a static spherically-symmetric system.  After solving \eqref{eq:OVgeneralised} for $p(r,t)$, which requires the imposition of a boundary condition on the pressure at some radius, one may complete the solution either by obtaining $f_1(r,t)$ from the $L_t\rho$ equation and hence $G(r,t)$ from any other equation that contains it, or by obtaining $G(r,t)$ from the $L_rp$ equation and then $f_1(r,t)$ from the $L_rf_1$ equation.

%We also note that the $L_t\rho$ equation may thus be regarded as the
%general relativistic equivalent of the continuity equation given in
%\eqref{eq:Newteqns}.

Finally, although the system of equations \eqref{eq:tetrels}--\eqref{eq:quantities} accommodates non-zero pressure, it is worth considering briefly the special case of a pressureless fluid.  In this case, the $L_rp$ equation forces $G$ to vanish, so the motion of the fluid particles becomes geodesic and the $L_rf_1$ equation forces $f_1=1$. Consequently, the components in the coordinate basis of the four-velocity of a fluid particle are $[v^\mu] \equiv [\dot{t},\dot{r},\dot{\theta},\dot{\phi}]=[1,g_2,0,0]$, where dots denote differentiation with respect to the particle's proper time $\tau$. Since $\dot{t}=1$, the coordinate time matches the proper time of all observers comoving with the fluid. Hence the Newtonian gauge is a synchronous one: a global `Newtonian' time is recovered on which all comoving observers agree (provided all clocks are synchronised initially)\footnote{In fact, these findings still hold in the slightly more general case in which there is no pressure {\em gradient}, thereby allowing for the fluid to have a non-zero homogeneous pressure.}.  Furthermore, combining the $L_tM$ equation and the definition of $M$ yields $(\partial_t + g_2\partial_r)g_2=-M/r^2 + \frac{1}{3}\Lambda r$, which has the form of the Euler equation in Newtonian fluid dynamics (recalling that $g_2$ is the fluid velocity $v$). Finally, setting $\Lambda=0$ for a moment, the definition of $M$ can itself be rearranged to give $\tfrac{1}{2}g_2-M/r = \frac{1}{2}(g_1^2-1)$, which is the Bernoulli equation for zero pressure and total (non-relativistic) energy per unit mass $\frac{1}{2}(g_1^2-1)$ (i.e.\ after subtraction of the rest-mass energy).

\subsection{Application to Schwarzschild spacetime}
\label{sec:tetradschwarzapp}

As an illustration of our approach, we now apply it to the special case in which the matter source is concentrated at the single point $r=0$ and the cosmological constant vanishes (see also \cite{Lasenby1998}).  For such a solution, $\rho=p=0$ everywhere away from the origin and so $L_rM=0$ and $L_tM=0$, which together imply $M=\mbox{constant}$. Retaining the symbol $M$ for this constant, one finds that the system of equations \eqref{eq:tetrels}--\eqref{eq:quantities} reduces just to the relationships \eqref{eq:tetrels} between the tetrad and spin-connection components and the definition of $M$ in \eqref{eq:quantities} (with $\Lambda \equiv 0$); no further equations yield new information.

We therefore have an under-determined system of equations and so some additional gauge-fixing is required to determine an explicit solution.  This occurs because in the final part of our gauge-fixing procedure described above, one chooses the timelike unit Lorentz frame vector at each point to coincide with the fluid four-velocity at that point, which clearly cannot be performed in a vacuum region. Nonetheless, one may instead choose the timelike unit frame vector to coincide with the four-velocity $\mathbf{u}$ of some radially-moving test particle (which need not necessarily be in free-fall), so that its components in the tetrad frame are $[\hat{u}^a] = [1,0,0,0]$ and hence in the coordinate basis one has $[u^\mu] \equiv [\dot{t},\dot{r},\dot{\theta},\dot{\phi}]=[f_1,g_2,0,0]$, as previously. This ensures that our previous physical interpretations of the tetrad and spin-connection components still hold. It remains, however, to choose a particular class of radially-moving test particle, and the simplest and most natural choice is a radially free-falling particle that was released from rest at $r=\infty$. From the definition \eqref{eq:quantities} of $M$ (with $\Lambda\equiv 0$), one sees that $g_1$ corresponds to the total energy per unit rest mass of an infalling particle, and so for a particle released from rest at infinity one should adopt the gauge condition $g_1=1$. It is then a simple matter to obtain expressions for the remaining tetrad components and spin-connection coefficients. The resulting non-zero tetrad components are
\begin{equation}
f_1=1,\qquad g_1=1, \qquad g_2=-\frac{2M}{r}, 
\label{eq:schwarztetrad}
\end{equation}
and the spin-connection coefficients $F$ and $G$ read
\begin{equation}
F=\left(\frac{2M}{r}\right)^{-1/2}\frac{M}{r^2},\qquad G=0.
\end{equation}
We note that the condition $G=0$ is clearly consistent with the geodesic motion of the test particles.

The line-element \eqref{eq:generalmetric} corresponding to the tetrad components \eqref{eq:schwarztetrad} is given by
\begin{equation}
ds^2 = dt^2 - \left(dr + \sqrt{\frac{2M}{r}}\,dt\right)^2 - r^2\,d\Omega^2,
\label{eq:pgmetric}
\end{equation}
which we recognise as the Schwarzschild spacetime line-element expressed in terms of Painlev\'e--Gullstrand coordinates \cite{painleve1921,gullstrand1922}.  This coordinate system has a number of desirable features. For example, the line-element is regular for all positive values of $r$ and the spacelike hypersurfaces $t=\mbox{constant}$ have Euclidean geometry. Moreover, from \eqref{eq:schwarztetrad}, the non-zero components of the four-velocity of a particle released from rest at infinity are immediately
\begin{equation}
\dot{t}=1,\qquad \dot{r}=-\sqrt{\frac{2M}{r}},
\end{equation}
and so we recover an essentially Newtonian description of the motion.  In particular, we see that $t$ coincides with the proper time of such particles. 

It is also of interest to consider briefly how to recover the standard form of the Schwarzschild line-element in Schwarzschild coordinates. This may be achieved by fixing the gauge by instead choosing the preferred class of test particle to have fixed spatial coordinates, in particular $\dot{r}=0$, which immediately requires $g_2=0$. It is then straightforward to obtain the remaining tetrad components and spin-connection coefficients. One thus finds
\begin{equation}
f_1=\left(1-\frac{2M}{r}\right)^{-1/2},\quad
g_1=\left(1-\frac{2M}{r}\right)^{1/2},\quad
g_2=0,
\label{eqn:schwarztetrad}
\end{equation}
and the spin-connection coefficients $F$ and $G$ are
\begin{equation}
F=0,\qquad G = \left(1-\frac{2M}{r}\right)^{-1/2}\frac{M}{r^2}.
\end{equation}
We note that $G\neq 0$ is consistent with non-geodesic motion of the test particles. Moreover, the corresponding line-element \eqref{eq:generalmetric} then takes the standard Schwarzschild form
\begin{equation}
ds^2 = {\displaystyle\left(1-\frac{2M}{r}\right)}\,dt^2
-\frac{dr^2}{{\displaystyle\left(1-\frac{2M}{r}\right)}}-r^2\,d\Omega^2.
\end{equation}
There exists a subtlety in the presence of a horizon, however, since it is not possible inside it for a test particle to remain at fixed spatial coordinates, and hence one cannot have $g_2=0$. This is discussed in detail in \cite{Lasenby1998}, where it is shown that the presence of a horizon is related to the onset of time-reversal asymmetry not easily identified using a wholly metric-based approach.

\subsection{Application to FRW spacetime}
\label{sec:tetradFRWapp}

As a second illustration of our approach, we apply it to the special case of a homogeneous and isotropic spacetime, as assumed in cosmology.  This corresponds to setting $\rho$ and $p$ to be functions of $t$ only. First we note that, unlike the Schwarzschild spacetime, there are no vacuum regions, and so no additional gauge-fixing will be required. It follows immediately from the $L_rp$ equation that $G=0$ and then the $L_rf_1$ equation implies $f_1=1$. Since $\rho$ is spatially uniform, one requires $M(r,t)=\tfrac{4}{3}\pi r^3 \rho$, which is consistent with the $L_rM$ equation. The $L_tM$ and $L_t\rho$ equations now jointly lead to the two conditions
\begin{align}
F&=\frac{g_2}{r}, \\
\partial_t\rho&=-\frac{3g_2(\rho+p)}{r},
\label{eq:rhoeqstep}
\end{align}
but the first condition can only be consistent with $L_rg_2$ equation if $F=H(t)$ and $g_2=rH(t)$. Finally, we obtain $g_1$ by first substituting the above expression for $M$ into \eqref{eq:quantities} and differentiating the result with respect to $r$ to obtain the equation $L_rg_1=(g_1^2-1)/r$, which may solved to yield $g_1^2=1+r^2\phi(t)$, where $\phi(t)$ is an arbitrary function of $t$. Then the $L_tg_1$ equation implies that $\partial_t\phi=-2H(t)\phi$.

Thus, gathering our results together, the non-zero tetrad components are given by
\begin{align}
f_1&=1,\nonumber \\
g^2_1&=1-kr^2\exp\left\{{\textstyle -2\int^t
  H(t')\,dt'}\right\},\nonumber \\
g_2 &= rH(t),
\label{eq:FRWtetrad}
\end{align}
where $k$ is an arbitrary constant of integration, and the spin-connection coefficients $F$ and $G$ are
\begin{equation}
F = H(t), \qquad G=0.
\end{equation}
We note that the condition $G=0$ demonstrates that the fluid particles move geodesically, since there are no pressure gradients. Moreover, on substituting the above results into \eqref{eq:rhoeqstep} and the expression \eqref{eq:quantities} for $M$, one obtains
\begin{align}
\partial_t\rho&=-3H(t)(\rho+p), \label{eq:tetradcosmo1}\\ 
\tfrac{8}{3}\pi\rho&=H^2(t)
-\tfrac{1}{3}\Lambda+k\exp\left\{{\textstyle -2\int^t
  H(t')\,dt'}\right\}.\label{eq:tetradcosmo2}
\end{align}
Finally, differentiating \eqref{eq:tetradcosmo2} with respect to $t$ and using \eqref{eq:tetradcosmo1}, one also obtains the further (although clearly not independent) dynamical equation
\begin{equation}
\partial_tH(t) + H^2(t)-\tfrac{1}{3}\Lambda = -\tfrac{4}{3}\pi(\rho+3p),
\label{eq:FRWaccel}
\end{equation}
which we recognise as the standard `acceleration' cosmological field equation expressed in terms of the Hubble parameter $H(t)$. Indeed, from \eqref{eq:FRWtetrad}, the non-zero components of the four-velocity of a fluid particle are immediately
\begin{equation}
\dot{t}=1,\qquad \dot{r} = rH(t),
\end{equation}
which both verifies that $t$ coincides with proper time of such particles and recovers Hubble's law.

Thus, our approach has led us to work directly with $H(t)$, which is an intrinsic and measurable quantity, rather than the more usual scale factor, which we will denote by $S(t)$. Nonetheless, we can make contact with the latter simply by setting $H(t) \equiv \partial_t S(t)/S(t)$, in which case $g_1^2 = 1-kr^2/S^2(t)$. The line-element \eqref{eq:generalmetric} corresponding to the tetrad components \eqref{eq:FRWtetrad} then reads
\begin{equation}
ds^2 = dt^2 -
\left(1-\frac{kr^2}{S^2(t)}\right)^{-1}\left[dr-rH(t)\,dt\right]^2-r^2\,d\Omega^2,
\label{eq:FRWmetricphys}
\end{equation}
and the dynamical equations \eqref{eq:tetradcosmo2} and \eqref{eq:FRWaccel} become
\begin{align}
\frac{[\partial_t S(t)]^2 + k}{S^2(t)} -\tfrac{1}{3}\Lambda &= \tfrac{8}{3}\pi\rho,\\
\frac{\partial^2_tS(t)}{S(t)}-\tfrac{1}{3}\Lambda &= -\tfrac{4}{3}\pi(\rho+3p),
\end{align}
which we recognise as Friedmann's cosmological field equations in their standard form.

The line-element \eqref{eq:FRWmetricphys} represents the FRW spacetime expressed in terms of a `physical' (i.e.\ non-comoving) radial coordinate $r$. We may rewrite this line-element in terms of a comoving radial coordinate $\hat{r} \equiv r/S(t)$, which yields the usual form
\begin{equation}
ds^2 = dt^2 - S^2(t)\left(\frac{d\hat{r}^2}{1-k\hat{r}^2}+\hat{r}^2\,d\Omega^2\right).
\end{equation}

\section{Comparison with LTB model}
\label{sec:ltb}

In contrast to our tetrad-based approach, the LTB model \cite{Lemaitre1933, Tolman1934, Bondi1947} is based on the use of a {\em comoving}\/ radial coordinate, which we denote by $\hat{r}$, and the assumption of a diagonal form for the metric. It is also usual to choose the time coordinate, which we denote by $\hat{t}$, to coincide with the proper time measured by observers comoving with the fluid, but for the moment we will consider a slightly more general version of the LTB metric in which this requirement is not enforced. Thus, we consider a line element of the form
\begin{equation}
  ds^2 = A^2d\hat{t}^2 - B^2d\hat{r}^2 - R^2 d\Omega^2,
  \label{eqn:metric_comov}
\end{equation}
where, in general, $A$, $B$ and $R$ may be arbitrary functions of both $\hat{r}$ and $\hat{t}$. Note that the standard form of the LTB metric corresponds to setting $A=1$.

We may understand the relationship between the line-element \eqref{eqn:metric_comov} and that given in \eqref{eq:generalmetric}, obtained using our tetrad-based approach, by performing a coordinate transformation that expresses the latter in terms of a comoving radial coordinate and brings it into diagonal form. We therefore consider the coordinate transformation
\begin{equation}
  t=\hat{t}, \quad r=r(\hat{r},\hat{t}), \quad \text{where} \quad
  \frac{\partial r}{\partial \hat{t}}=\frac{g_2}{f_1}.
  \label{eqn:transf}
\end{equation}
Note that, although the time coordinates coincide, we still label the new one as $\hat{t}$, since the partial derivatives $\partial/\partial t$ and $\partial/\partial\hat{t}$ will, in general, be different because they hold fixed $r$ and $\hat{r}$, respectively.  One may verify that $\hat{r}$ is a comoving radial coordinate by recalling that the four-velocity components of a comoving observer in the Newtonian gauge are $[v^{\mu}]=(\dot{t}, \dot{r},\dot{\theta}, \dot{\phi}) = (f_1, g_2, 0,0)$, which transform under \eqref{eqn:transf} into $[\hat{v}^{\mu}] =(\dot{\hat{t}},\dot{\hat{r}},\dot{\theta},\dot{\phi}) = (f_1,0,0,0)$.  The physical nature of the transformation \eqref{eqn:transf} may be further clarified by noting that
\begin{equation}
        \frac{\partial}{\partial \hat{t}} = \frac{\partial t}{\partial \hat{t}}  \frac{\partial}{\partial t} + \frac{\partial r}{\partial \hat{t}}  \frac{\partial}{\partial r}   
        = \frac{1}{f_1} \left( f_1 \frac{\partial}{\partial t} + g_2 \frac{\partial}{\partial r}\right)   
        = \frac{1}{f_1} L_t,
    \label{eqn:d_dt_comov}
\end{equation}
where $L_t$, defined in \eqref{eq:operators}, is the derivative with respect to the proper time of a comoving observer; indeed this is consistent with our finding above that $\dot{\hat{t}}=f_1$. Since $L_t$ may be considered as a relativistic form of convective derivative, one may interpret the transformation \eqref{eqn:transf} as moving from a Eulerian to a Lagrangian description of the fluid motion. Similarly, one finds that
\begin{equation}
  \frac{\partial}{\partial \hat{r}} = \frac{1}{g_1} \frac{\partial r}{\partial \hat{r}} L_r,
  \label{eqn:d_dr_comov}
\end{equation}
where $L_r$, also defined in \eqref{eq:operators}, is the derivative with respect to the proper radial distance of a comoving observer.

Under the transformation \eqref{eqn:transf} to a comoving radial coordinate, the line-element \eqref{eq:generalmetric} takes the diagonal form
\begin{equation}
  ds^2= \frac{1}{f_1^2}\,d\hat{t}^2 - \frac{1}{g_1^2} \left(
  \frac{\partial r}{\partial \hat{r}} \right)^2 d\hat{r}^2 - r^2
  d\Omega^2.
  \label{eqn:met_trans}
\end{equation}
One should first note that this has been achieved without having to specify $\partial r/\partial \hat{r}$; this demonstrates that the (generalised) LTB metric \eqref{eqn:metric_comov} possesses a residual gauge freedom, in contrast to the line-element \eqref{eq:generalmetric} (recall that the final gauge choice made in Section~\ref{sec:framework} leaves the form of \eqref{eq:generalmetric} unchanged).

Comparing \eqref{eqn:metric_comov} and \eqref{eqn:met_trans}, one first identifies that $r=R(\hat{r},\hat{t})$ and hence the three non-zero tetrad components used in Section~\ref{sec:framework} are given by
\begin{equation}
f_1 = \frac{1}{A}, \qquad 
g_1 = \frac{1}{B}\partial_{\hat{r}}R, \qquad
g_2 = \frac{1}{A}\partial_{\hat{t}}R,
%f_1 = \frac{1}{A}, \qquad 
%g_1 = \frac{1}{B}\frac{\partial R}{\partial \hat{r}}, \qquad
%g_2 = \frac{1}{A}\frac{\partial R}{\partial \hat{t}},
\label{eqn:tetrads_comov}
\end{equation}
where the final result is obtained using \eqref{eqn:transf}.  The expressions for the spin-connection coefficients $F$ and $G$ are obtained from the relations (\ref{eq:tetrels}) and read
\begin{equation}
F = \frac{1}{AB}\partial_{\hat{t}}B, \qquad
G = \frac{1}{AB}\partial_{\hat{r}}A.
%F = \frac{1}{AB}\frac{\partial B}{\partial \hat{t}}, \qquad
%G = \frac{1}{AB}\frac{\partial A}{\partial \hat{r}}.
\label{eq:LTBspinconn}
\end{equation}

Substituting the expressions \eqref{eqn:tetrads_comov} and \eqref{eq:LTBspinconn} into the dynamical and continuity equations \eqref{eq:alleqns}, one then obtains
\begin{align}
\partial_{\hat{r}}p &= -\frac{(\rho + p)}{A}\partial_{\hat{r}}A,
\label{eqn:dpdr_comov}\\
\partial_{\hat{r}}M &=4 \pi R^2 \rho \,\partial_{\hat{r}}R, \label{eqn:LrM_comov}\\
\partial_{\hat{t}}\rho & = -\left( \rho + p \right)
\left( \frac{2}{R}\partial_{\hat{t}}R +
\frac{1}{B} \partial_{\hat{t}}B \right), \label{eqn:cont}\\
\partial_{\hat{t}}M &=-4 \pi R^2 p \,\partial_{\hat{t}}R, \label{eqn:LtM_comov}
%\frac{\partial p}{\partial \hat{r}} 
%&= -\frac{1}{A}\frac{\partial A}{\partial \hat{r}} (\rho + p),
%\label{eqn:dpdr_comov}\\
%\frac{\partial M}{\partial \hat{r}} &=4 \pi R^2 \frac{\partial
%  R}{\partial \hat{r}} \rho, \label{eqn:LrM_comov}\\
%\frac{\partial \rho}{\partial \hat{t}} & = -\left( \rho + p \right)
%\left( 2\frac{1}{R}  \frac{\partial R }{\partial \hat{t}} +
%\frac{1}{B} \frac{\partial B }{\partial \hat{t}} \right), \label{eqn:cont}\\
%\frac{\partial M}{\partial \hat{t}} &=-4 \pi R^2 \frac{\partial
%  R}{\partial \hat{t}} p, \label{eqn:LtM_comov}
\end{align}
where the expression for $M$ in \eqref{eq:quantities} now becomes
\begin{equation}
  \frac{2M}{R} = \frac{1}{A^2}\left(\partial_{\hat{t}}R\right)^2-
  \frac{1}{B^2}\left(\partial_{\hat{r}}R \right)^2 +1-\tfrac{1}{3}\Lambda R^2.
  \label{eqn:M_comov}
\end{equation}

The assumed form \eqref{eqn:metric_comov} of the line-element and the system of equations \eqref{eqn:dpdr_comov}--\eqref{eqn:M_comov} constitute a generalised form of the LTB model that can accommodate pressure and a non-zero cosmological constant. Nonetheless, unlike the tetrad-based approach, this model still possesses a gauge freedom, since $\partial_{\hat{r}}R$ remains arbitrary. Thus, to determine the evolution of the system, one must first choose a form for the function $R(\hat{r},\hat{t}_*)$ at some time $\hat{t}_*$ (usually given as an initial condition), which is not easily interpreted physically.

To make contact with the standard LTB model, we may now set $A=1$ in the line-element \eqref{eqn:metric_comov}, so that $\hat{t}$ coincides with the proper time measured by observers comoving with the fluid. Hence, like the Newtonian gauge, the standard LTB model employs a synchronous coordinate system.  In terms of the tetrad components \eqref{eqn:tetrads_comov} and spin-connection coefficients \eqref{eq:LTBspinconn}, setting $A=1$ corresponds to setting $f_1=1$ and $G=0$, and hence \eqref{eqn:d_dt_comov} shows that the operators $\partial_{\hat{t}}$ and $L_t$ then coincide. Thus, from the $L_tg_1$ equation in \eqref{eq:tetrels} one finds that $g_1=g_1(\hat{r})$ is a function only of $\hat{r}$. Using the expression for $g_1$ in \eqref{eqn:tetrads_comov} and adopting the standard notation used in LTB models, we therefore define the function $E(\hat{r})$ by
\begin{equation}
1+2E(\hat{r}) \equiv \frac{1}{B^2}(\partial_{\hat{r}}R)^2,
\label{eq:erdef}
\end{equation}
which we may choose arbitrarily provided that $E(\hat{r}) > -\tfrac{1}{2}$.  It is also immediately clear from \eqref{eqn:dpdr_comov} that setting $A=1$ requires the pressure gradient to vanish. Thus, the standard LTB line-element can at best accommodate a fluid with a non-zero homogeneous pressure. It is usual, however, to assume simply that the fluid is pressureless, in which case \eqref{eqn:LtM_comov} shows that $M=M(\hat{r})$ is a function only of $\hat{r}$.  Given our earlier interpretation of $M$ in the tetrad-based approach, one may thus verify the usual interpretation of $M(\hat{r})$ in the LTB model as the mass-energy interior to a sphere of comoving radial coordinate $\hat{r}$, which is naturally time-independent in the absence of pressure.

Collecting together our results, setting $A=1$ in \eqref{eqn:metric_comov} and assuming a pressureless fluid thus imply that equations \eqref{eq:erdef}, \eqref{eqn:M_comov} and \eqref{eqn:LrM_comov} may be written, respectively, as
\begin{align}
B^2 & = \frac{(\partial_{\hat{r}}R)^2}{1+2E(\hat{r})},  \label{eq:LTB1}\\
E(\hat{r}) & = \tfrac{1}{2} (\partial_{\hat{t}}R)^2 -
\frac{M(\hat{r})}{R} -\tfrac{1}{6}\Lambda R^2, \label{eq:LTB2} \\
4\pi\rho& = \frac{\partial_{\hat{r}}M(\hat{r})}{R^2\partial_{\hat{r}}R},
\label{eq:LTB3}
\end{align}
which we recognise as the standard equations for the LTB model (with the inclusion of a non-zero cosmological constant), where $M(\hat{r})$ and $E(\hat{r})$ may be chosen arbitrarily (subject to $M(\hat{r}) > 0$ and $E(\hat{r}) > -\tfrac{1}{2}$). Indeed, since the first equation in the set may be written as $E(\hat{r})=\tfrac{1}{2}[g_1(\hat{r})-1]$, our approach also allows us to verify the usual interpretation of $E(\hat{r})$ as the total (non-relativistic) energy per unit rest mass of a fluid particle (i.e.\ after subtracting its rest mass energy) at comoving coordinate radius $\hat{r}$. In addition to the usual set of equations \eqref{eq:LTB1}--\eqref{eq:LTB3}, our approach also naturally yields the continuity equation \eqref{eqn:cont}, which originates from the contracted Bianchi identities, and in this case reduces to
\begin{equation}
\partial_{\hat{t}}\rho = - \left(
\frac{2}{R}\partial_{\hat{t}}R
+\frac{1}{\partial_{\hat{r}}R}\partial_{\hat{t}}\partial_{\hat{r}}R\right)\rho.
\label{eq:LTBcont}
\end{equation}

Finally, it is also worth mentioning that on integrating \eqref{eq:LTB2}, it is usual to introduce another arbitrary function $t_{\rm b}(\hat{r})$, known as the ``bang-time'' and interpreted as a Big-Bang singularity surface at which $R$ vanishes, i.e.\ $R(\hat{r},\hat{t}_{\rm b}(\hat{r}))=0$. From our tetrad-based approach, however, we see that this condition corresponds simply to identifying the origin $r=0$ of our `physical' radial coordinate.

\subsection{Application to Schwarzschild spacetime}

As an illustration of the LTB model, and in particular to compare it with the tetrad-based approach, we now apply it to the same physical situation as we considered in Section~\ref{sec:tetradschwarzapp}, namely that of a matter source concentrated into a single point and a vanishing cosmological constant. As previously, for such a solution, $\rho=p=0$ everywhere away from the point mass, and so \eqref{eq:LTB3} implies that $M=\mbox{constant}$. Once again, the remaining system of equations is under-determined, and so some gauge-fixing is required. First we must choose a form for the arbitrary function $E(\hat{r})$. Similar to the approach adopted in Section~\ref{sec:tetradschwarzapp}, we may base our choice on some class of radially-moving test particle and, once again, the most natural choice is a radially free-falling particle released from rest at infinity. From the definition \eqref{eq:erdef}, we see that the choice of $g_1=1$ in our tetrad-based approach is equivalent to setting $E(\hat{r})=0$, which corresponds to the particle having zero energy (after subtraction of its rest mass energy).

Thus, with $M(\hat{r})=M$ (a constant) and $E(\hat{r})=0$ (and $\Lambda=0$), the LTB equation \eqref{eq:LTB2} reduces to
\begin{equation}
\partial_{\hat{t}}R = -\sqrt{\frac{2M}{R}},
\end{equation}
where we have taken the negative square root since the test particle is radially infalling. This equation may be immediately integrated to give $\tfrac{2}{3}R^{3/2}=-(2M)^{1/2}[\hat{t}-\hat{t}_b(\hat{r})]$, where the ``bang-time'' $\hat{t}_{\rm b}(\hat{r})$ may be an arbitrary function of $\hat{r}$, but is usually chosen such that $R(\hat{r},\hat{t}_{\rm b}(\hat{r}))=0$, which in this case requires $\hat{t}_{\rm b}(\hat{r})=\hat{r}$. Thus, after this additional gauge-fixing, which was not required in the tetrad based approach, one has
\begin{equation}
R=\left[\frac{9M}{2}(\hat{r}-\hat{t})^2\right]^{1/3},\qquad
B=\tfrac{2}{3}\left[\frac{9M}{2(\hat{r}-\hat{t})}\right]^{1/3}
\end{equation}
where the second result is derived straightforwardly from \eqref{eq:LTB1}. Substituting these expressions into \eqref{eqn:metric_comov} with $A=1$, one immediately obtains
\begin{equation}
ds^2 = d\hat{t}^2 - 
\tfrac{4}{9}\left[\frac{9M}{2(\hat{r}-\hat{t})}\right]^{2/3}\,d\hat{r}^2-
\left[\frac{9M}{2}(\hat{r}-\hat{t})^2\right]^{2/3}\,d\Omega^2,
\label{eq:lemaitremetric}
\end{equation}
which is the line-element for the Schwarzschild spacetime expressed in Lema\^{i}tre coordinates \cite{Lemaitre1933}. This line-element is regular for all values of $\hat{r}$, except at $\hat{r}-\hat{t}=0$, which corresponds to the location of the point mass (i.e.\ at $r=0$ in terms of the `physical' radial coordinate used in Section~\ref{sec:tetradschwarzapp}). It is straightforward to show that radially free-falling test particles released from rest at infinity have fixed values of $\hat{r}$, $\theta$ and $\phi$, which therefore constitute comoving coordinates for such particles. From \eqref{eq:lemaitremetric}, one thus sees that $\hat{t}$ does indeed coincide, by construction, with the proper time of such particles.

It is interesting that, although the tetrad-based approach and the LTB model both employ synchronous time coordinates and are based on the trajectories of radially infalling particles released from rest at infinity, the two methods naturally lead to the very different line-elements \eqref{eq:pgmetric} and \eqref{eq:lemaitremetric}. This occurs because of the use of a `physical' radial coordinate in the former, whereas the latter employs a comoving radial coordinate, and also the requirement that the LTB line-element be diagonal. In the authors' opinion, the former line-element, expressed in Painlev\'e--Gullstrand coordinates, is the more easily interpreted physically.

\subsection{Application to FRW spacetime}
\label{sec:LTBFRWapp}

As an second illustration of the LTB model, we now apply it to the special case of a homogeneous and isotropic spacetime, as considered in Section~\ref{sec:tetradFRWapp} using the tetrad-based approach. As before, this corresponds to setting $\rho$ to be a function of $\hat{t}$ only, but for the LTB model we are limited to considering only pressureless fluids and so $p=0$. In contrast to the tetrad-based approach, one must begin by making a gauge choice for the form for $M(\hat{r})$. This is most naturally achieved by introducing the scale factor $S(\hat{t})$ at the outset, such that $\rho(\hat{t})=\rho_0[S_0/S(\hat{t})]^3$, where $\rho_0 \equiv \rho(\hat{t}_0)$ and $S_0 \equiv S(\hat{t}_0)$ are defined at some cosmic time $\hat{t}=\hat{t}_0$, usually taken in cosmology to be the current epoch. Keeping in mind the physical interpretation of $M(\hat{r})$, it is then simplest to assume the form $M(\hat{r})=\tfrac{4}{3}\pi\rho_0S_0^3\hat{r}^3$. It then remains to make a further gauge choice for the form of $E(\hat{r})$. Despite its physical interpretation as the total (non-relativistic) energy per unit rest mass of a fluid particle with comoving radial coordinate $\hat{r}$, it is not immediately clear how to assign it a functional form. Nonetheless, the form of $g_1$ obtained in the tetrad based approach in Section~\ref{sec:tetradFRWapp} suggests setting $E(\hat{r})=-\tfrac{1}{2}k\hat{r}^2$, where $k$ is a constant, which one might be persuaded {\em post-hoc}\/ is consistent with the above physical interpretation.

Thus, with $M(\hat{r})=\tfrac{4}{3}\pi\rho_0S_0^3\hat{r}^3$ and $E(\hat{r})=-\tfrac{1}{2}k\hat{r}^2$, the LTB equations \eqref{eq:LTB1} and \eqref{eq:LTB3} lead straightforwardly to
\begin{equation}
R = S(\hat{t})\hat{r}, \qquad B = \frac{S(\hat{t})}{\sqrt{1-k\hat{r}^2}},
\end{equation}
and the corresponding line-element (\ref{eqn:metric_comov}) (with $A=1$) takes the usual FRW form
\begin{equation}
ds^2 = dt^2 - S^2(t)\left(\frac{d\hat{r}^2}{1-k\hat{r}^2}+\hat{r}^2\,d\Omega^2\right).
\end{equation}
Moreover, the remaining LTB equations \eqref{eq:LTB2} and \eqref{eq:LTBcont} then yield the standard Friedmann equation and cosmological continuity equation, respectively, namely
\begin{align}
\frac{[\partial_{\hat{t}} S(\hat{t})]^2 + k}{S^2(\hat{t})} -\tfrac{1}{3}\Lambda &=
\tfrac{8}{3}\pi\rho, \\
\partial_{\hat{t}}\rho&=-3H(\hat{t})\rho,
\end{align}
where we have defined the Hubble parameter $H(\hat{t})\equiv \partial_{\hat{t}}S(\hat{t})/S(\hat{t})$. These two equations can be combined to yield the standard `acceleration' cosmological field equation for zero pressure, if desired.

Thus, we see that the LTB model has led us directly to working in terms of the scale factor, in contrast to the tetrad-based approach used in Section~\ref{sec:tetradFRWapp}, which led naturally to the Hubble parameter, which is a directly measurable quantity. Again, this difference results from the use of a comoving radial coordinate in the LTB model and the assumption of a diagonal metric, as compared to using a `physical' radial coordinate in the tetrad-based approach and making no such restriction on the form of the metric. Moreover, considerable gauge-fixing was required in the LTB model to obtain a definite form for the solution, whereas this was unnecessary in the tetrad-based approach.

\section{Generalised Swiss cheese model}
\label{sec:gscm}

We now apply our tetrad-based approach to a generalised form of the Swiss cheese model. In its classic form, the Swiss cheese model consists of an exterior expanding FRW universe, albeit pressureless, in which a uniform pressureless spherical object is embedded and surrounded by a `compensating void' that itself expands into the background and ensures that there is no net gravitational effect on the exterior universe. Such models were employed in some of the earliest attempts to describe non-linear cosmological inhomogeneities \cite{Rees1968,Nottale1982}, since they have the advantage that analytical calculations can be performed, and compensation ensures observations in the exterior region can be modelled unambiguously. Nonetheless, the matter and velocity distributions are clearly unrealistic.

As mentioned in the Introduction, more realistic models can be constructed by working with continuous density and velocity profiles, while still restricting to spherical symmetry and ignoring pressure. Previous work using LTB models \cite{Garcia-bellido2008,Alonso2010} has usually ignored compensation, and this can lead to subtleties in modelling observations in the exterior region, since it is not described by a homogeneous FRW cosmology. Constructing compensated models can be difficult, however, since the initial density and velocity profiles must be carefully chosen so that streamline crossing is avoided. Otherwise, shock fronts form and one must include pressure to produce a realistic model.  For example, if one perturbs just the density profile without perturbing the velocity profile, then streamline crossing is inevitable in any compensated model other than those of Swiss cheese type (where the problem would occur only in the vacuum region). In \cite{Lasenby1999,Dabrowski1999}, we therefore presented a family of continuous initial density and velocity profiles, described by polynomials in radius, that avoid the problem of streamline crossing in a very simple manner, while keeping the density profile compensated and realistic, and which are consistent with having grown from primordial fluctuations in the very early universe. The evolution of the resulting spatially-finite inhomogeneity was then determined in the absence of pressure using our tetrad-based method.

In this Section, however, our primary focus is not the modelling of realistic cosmological inhomogeneities or the prediction of observational effects in the exterior region. Rather, we wish merely to extend the analysis presented in \cite{NLH3} (hereinafter NLH3) by applying our tetrad-based method to a generalised form of Swiss cheese model, in which we allow the fluid in the interior and exterior regions to support pressure, in general, and do not demand that the interior region be compensated.  Aside from intellectual curiosity, the motivations for this study are two-fold: we first wish to verify that Birkhoff's theorem holds in the vacuum region, the usual interpretation of which has recently been brought into question for related systems \cite{Zhang2012}; and, second, we wish to consider the validity of the theoretical arguments that underpin the $R_{\rm h}=ct$ cosmological model \cite{Melia2007,Melia2009,MeliaShevchuk2012}, which has recently received considerable attention \cite{vanOirschot2010,LewisvanOirschot2012,Mitra2014}. These investigations are presented, respectively, in Sections~\ref{sec:birkhoff} and~\ref{sec:rctmodel} below.

\subsection{Model specification}

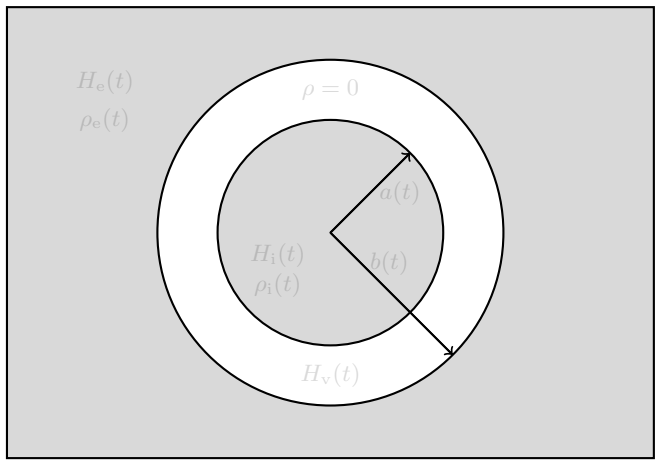
\begin{figure}[t]
%  \centering
  \begin{tikzpicture}[thick,fill opacity=.15,draw opacity=1, text opacity=1]
    \fill[black]  (-4.3,-3) rectangle (4.3cm,3cm);  
    \draw[name path=rec,thick] (-4.3,-3) rectangle (4.3cm,3cm); 
    \fill[white, opacity=1] (0cm,0cm) circle(2.3cm); 
    \draw[name path=circle2, thick] (0cm,0cm) circle(2.3cm);
    \filldraw[fill=black, draw=black] (0,0) circle(1.5cm); 
    \draw[thick,->] (0,0) -- (1.06066,1.06066) node[midway,below,right]{$a(t)$};
    \draw[thick,->] (0,0) -- (1.626346,-1.626346) node[near start,above,right]{$b(t)$};
    \node at (0, 1.9) {$\rho=0$} ;
    \node at (0, -1.9) {$H_{\rm v}(t)$} ;
    \node at (-0.7,-0.7) {$\rho_{\rm i}(t)$};
    \node at (-0.7,-0.3) {$H_{\rm i}(t)$};
    \node at (-3, 1.5) {$\rho_{\rm e}(t)$};
    \node at (-3, 2) {$H_{\rm e}(t)$};
  \end{tikzpicture}
  \caption{Generalised Swiss cheese model: a spherical interior region
    of uniform density $\rho_{\rm i}(t)$ and radius $a(t)$ is
    surrounded by a vacuum region of radius $b(t)$, which itself
    resides in an exterior region with uniform density $\rho_{\rm
      e}(t)$. The rates of expansion of the interior, vacuum and
    exterior regions are characterized by the `Hubble parameters'
    $H_{\rm i}(t)$, $H_{\rm v}(t)$ and $H_{\rm e}(t)$,
    respectively. In general, the fluid can support pressure and the
    interior region need not be compensated.\label{fig:model_diag}}
\end{figure}

The generalised Swiss cheese model is illustrated in Fig.~\ref{fig:model_diag} and consists of a spherical interior region of uniform density $\rho_{\rm i}(t)$ and radius $a(t)$ surrounded by a vacuum region of radius $b(t)$, which itself resides in an exterior region with uniform density $\rho_{\rm e}(t)$. The rates of expansion of the interior, vacuum and exterior regions are characterized by the `Hubble parameters' $H_{\rm i}(t)$, $H_{\rm v}(t)$ and $H_{\rm e}(t)$, respectively (the definition of $H_{\rm v}(t)$ is discussed below). In general, the interior region need not be compensated and the fluid in both the interior and exterior
regions can support pressure.  

From the figure, we may write down an expression for the total mass-energy $M(r,t)$ contained within a sphere of physical radius $r$ at time $t$. It is clear that
\begin{equation}
  M=
    \begin{cases}
      \frac{4}{3} \pi \rho_{\rm i}(t)r^3, &  r\leq a(t),\\ 
      \frac{4}{3} \pi \rho_{\rm i}(t)a(t)^3 \equiv M_0, &  a(t) \le r\le b(t),\\ 
      \frac{4}{3} \pi \rho_{\rm e}(t)r^3 + m(t), &  r \geq b(t),
    \end{cases}
\label{eq:Mdefns}
\end{equation}
where $M_0$ is a constant and $m(t)= M_0 - \frac{4}{3} \pi \rho_{\rm e}(t) b(t)^3$ is the mass contained within $b(t)$ at time $t$, {\em in excess}\/ of that which would be present due to the exterior background alone.  For a compensated interior region, one thus has $m(t) \equiv 0$. We note that the system considered in NLH3 corresponds to setting $b(t)=a(t)$, so that there is no vacuum region.

As will we show below, in the general case where the fluid supports pressure, to determine the dynamical evolution of the system completely one must specify the internal and external Hubble parameters $H_{\rm i}(t)$ and $H_{\rm e}(t)$, together with the evolution $a(t)$ and $b(t)$ of the two boundaries (and the density $\rho_{\ast}\equiv \rho_{\rm i}(t_\ast)$ of the interior region at some reference time $t=t_\ast$). Typically, one should take $H_{\rm e}(t)$ to correspond to some expanding exterior universe of interest, but $H_{\rm i}(t)$, $a(t)$ and $b(t)$ can, in principle, have any form.\footnote{As shown by NLH3, however, in the case where $b(t)=a(t)$, so that there is no vacuum region, the evolution $a(t)$ of the single boundary between the two fluid regions cannot be set arbitrarily, but is instead determined by specifying $H_{\rm i}(t)$ and $H_{\rm e}(t)$ (together with the interior density $\rho_{\ast}\equiv \rho_{\rm i}(t_\ast)$ and the radius $a_\ast\equiv a(t_\ast)$ at some reference time $t=t_\ast$).}  This follows both from the presence of the vacuum region and from allowing the relationship between the fluid pressure and density to be arbitrary, since then the interplay between pressure and gravity may allow expansion or contraction of the interior and vacuum regions at any rate. This freedom would disappear, however, if one imposed an equation of state on the fluid. In particular, in the special case of a pressureless fluid, $a(t)$ is straightforwardly determined from $H_{\rm i}(t)$ (and the radius $a_\ast\equiv a(t_\ast)$ at some reference time $t=t_\ast$).

Following our approach in NLH3, in the interior and exterior regions, we assume a single `phenomenological' fluid.  This avoids the complexity of an explicit multi-fluid treatment, whereby one would separate the fluid in each region into its baryonic and dark matter components.  In particular, in each region we assume a single overall (uniform) fluid density and a single associated (effective) pressure. It is envisaged that the pressure comprises the ordinary gas pressure due to baryonic matter, and an effective pressure from the dark matter that arises from the motions of dark matter particles having undergone phase-mixing and relaxation (see \cite{Lynden-Bell1967} and \cite{Binney2008}).

\subsection{Boundary conditions}
\label{sec:bcs}

Any spatial surface at which the density is discontinuous, and which may in general be moving, will trace out a 3-dimensional (timelike) hypersurface $\Sigma$ in spacetime on which the solution must satisfy the Israel junction conditions \citep{Israel1966,Israel1967}. If $\hat{n}_\mu$ are the covariant components of the unit (spacelike) normal to $\Sigma$, pointing from the inside to the outside, then the Israel junction conditions require continuity both of the induced metric $h_{\mu\nu}=g_{\mu\nu}+\hat{n}_\mu\hat{n}_\mu$ and the extrinsic curvature $K_{\alpha\beta} = {h_\alpha}^\mu {h_\beta}^\nu \nabla_\mu \hat{n}_\nu$ on $\Sigma$.

For the model illustrated in Fig.~\ref{fig:model_diag}, two such hypersurfaces are defined by $\Sigma(t,r) \equiv r-x(t) = 0$, where $x(t)$ can equal either $a(t)$ or $b(t)$. As discussed in \cite{NLH3}, the components $\hat{n}_\mu$ are given by
\begin{equation}
[\hat{n}_\mu] = \frac{[-\partial_t x, 1, 0, 0]}
{|f_1^2 \partial_t x-2f_1 g_1 \partial_t x + g_2^2-g_1^2|^{1/2}},
\label{eq:nhatgen}
\end{equation}
where $\partial_t x \equiv dx(t)/dt$; one may readily verify that $\hat{n}_\mu \hat{n}^\mu = -1$, as required. One sees immediately that, for the induced metric to be continuous across the boundary $x(t)$, one requires all three non-zero tetrad components $f_1$, $g_1$ and $g_2$ to be continuous there.

Recalling that $g_2$ is the rate of change of the $r$ coordinate of a fluid particle with respect to its proper time, and can thus be considered as the fluid velocity, the physical interpretation of the continuity of $g_2$ is that matter does not cross the boundary $x(t)$ in either direction.  This is consistent with the boundary $x(t)$ comoving with the fluid, so that the situation depicted in Fig.~\ref{fig:model_diag} does indeed hold at all times. It does {\em not}\/ imply, however, that $M(x(t), t)$ is constant, since this quantity denotes the total energy within $x(t)$, which may change as $x(t)$ evolves, as is clear from the $L_tM$ and $L_rM$ equations in (\ref{eq:alleqns}). Moreover, since $L_t$ corresponds to the derivative with respect to the proper time of an observer comoving with the fluid, then one requires
\begin{equation}
  L_t x(t) = g_2(x(t),t),
  \label{eqn:a_dash}
\end{equation}
where $L_t$ is evaluated at the boundary $x(t)$. Thus, on the hypersurface $\Sigma$, one has $dx(t)/dt = g_2/f_1$ and the expression (\ref{eq:nhatgen}) simplifies to
\begin{equation}
[\hat{n}_\mu] = \frac{1}{g_1}[-g_2/f_1, 1, 0, 0].
\label{eq:nhatdown}
\end{equation}
After a long but straightforward calculation, one then finds that the only non-zero components of the extrinsic curvature of $\Sigma$ are\footnote{The expression for $K_{00}$ given here differs from that in \cite{NLH3}, since the latter is incorrect owing to a sign error in the original calculation. Nonetheless, both forms lead to the same conclusion regarding the continuity of $\partial_rf_1$ at the boundary.}
\begin{equation}
K_{00} = \frac{g_1}{f_1^3}\partial_r f_1,\qquad
K_{11} = g_1r,\qquad
K_{22} = g_1r \sin^2\theta.
\end{equation}
Since the first Israel junction condition requires $f_1$, $g_1$ and $g_2$ all to be continuous at the boundary $x(t)$, then the second junction condition requires only that, in addition, $\partial_r f_1$ is continuous there.

The above junction conditions have consequences for the continuity of other variables of interest. In particular, from the $L_rf_1$ equation in \eqref{eq:tetrels}, one has $G = -(g_1/f_1)\partial_r f_1$, which must therefore also be continuous at the boundary. Moreover, the $L_rp$ equation in \eqref{eq:alleqns} and the continuity of $g_1$ and $G$ imply that the pressure $p$ is also continuous across the boundary, although its radial derivative, in general, has a step there.

Finally, we also adopt the boundary condition at large $r$ that all physical quantities tend to those of the exterior cosmology. For spatially-flat and open universes, this corresponds to the limit $r\to \infty$, whereas for a closed universe one must instead consider the limit $x(t) \ll r < S(t)$, where $S(t)$ is the universal scale factor which also corresponds to the curvature scale for a closed universe. In each case, we require the line-element \eqref{eq:generalmetric} to tend at large $r$ to the corresponding FRW line-element \eqref{eq:FRWmetricphys} with $H(t) = H_{\rm e}(t)$.  Thus, for large $r$, one requires
%\begin{align}
%f_1 &\to 1,\nonumber \\
%g_1 &\to \sqrt{1-\frac{kr^2}{S^2(t)}},\nonumber\\
%g_2 &\to rH_{\rm e}(t).\label{eq:bclarger}
%\end{align}
\begin{equation}
f_1 \to 1,\qquad
g_1 \to \sqrt{1-\frac{kr^2}{S^2(t)}},\qquad
g_2 \to rH_{\rm e}(t).\label{eq:bclarger}
\end{equation}

\subsection{Interior and exterior regions}

From \eqref{eq:Mdefns}, one sees that the forms for $M$ in the interior and exterior regions are the same as the model considered in NLH3, albeit with a different definition of $m(t)$. Moreover, the same boundary conditions \eqref{eq:bclarger} apply at large $r$. Therefore, many of the equations derived in NLH3 remain valid.

Specifically, in the interior region, the non-zero tetrad components are again given by
\begin{align}
	f_{1,{\rm i}} &= -\frac{3H_{\rm i}(t)\left(\rho_{\rm
            i}(t)+p_{\rm i} \right)}{\rho_{\rm i}'(t)}, \label{eqn:f1_int}\\
	g_{1,{\rm i}} &= \sqrt{1+r^{2} \left( H_{\rm
            i}^{2}(t)-\tfrac{8 }{3} \pi 
\rho_{\rm i}(t)- \tfrac{1}{3} \Lambda \right) }, \label{eqn:g1_int}\\
    g_{2,{\rm i}}  &= rH_{\rm i}(t),\label{eqn:g2_int}
\end{align}
where, following NLH3, a prime denotes differentiation with respect to $t$. In order to evaluate the above expressions for $f_{1,{\rm i}}$ and $g_{1,{\rm i}}$, one requires forms for $\rho_{\rm i}(t)$ and $p_{\rm i}$.  Substituting the above expression for the fluid velocity $g_{2,{\rm i}}$ and the enclosed mass $M=\tfrac{4\pi}{3}\rho_{\rm i}(t)r^3$ from \eqref{eq:Mdefns} into the generalised Oppenheimer--Volkov equation (\ref{eq:OVgeneralised}), and integrating, will yield an (integral) expression for $p_{\rm i}$ in terms of $H_{\rm i}(t)$ and $\rho_{\rm i}(t)$, after imposing the condition that the pressure is continuous across the boundary $a(t)$ and hence vanishes there.  Thus, it only remains to determine $\rho_{\rm i}(t)$, which is straightforwardly obtained for a given boundary evolution $a(t)$ by recalling that $M_0 \equiv \frac{4\pi}{3}\rho_{\rm i}(t)a^3(t)$ is a constant. In the special case of a pressureless fluid, it is worth noting that one immediately has $f_{1,{\rm i}}=1$ and so \eqref{eqn:f1_int} can be integrated to obtain $\rho_{\rm i}(t)$, which then determines $a(t)$.

For the exterior region, the non-zero tetrad elements are given by
\begin{widetext}
\begin{align}
    f_{1,{\rm e}} &=-\frac{3H_{\rm e}(t)\left(\rho_{\rm e}(t)+p_{\rm e}\right)}{\rho_{\rm e}'(t)},\label{eqn:f1_ext}\\
    g_{1,{\rm e}} &=\sqrt{1- \frac{2m(t)}{r} + \left( H_{\rm e}(t)^{2}-\tfrac{8}{3} \pi \rho_{\rm e}(t)-\tfrac{1}{3}\Lambda \right) r^2 + \frac{9H_{\rm e}^2(t)m'(t)}{16\pi^2 \rho_{\rm e}'^2(t)r^4} \left( \tfrac{8}{3}\pi \rho_{\rm e}'(t) r^3 + m'(t) \right)}, \label{eqn:g1_ext}\\
    g_{2,{\rm e}} &=r H_{\rm e}(t) + \frac{3m'(t)H_{\rm e}(t)}{4 \pi r^{2}\rho_{\rm e}'(t)}.\label{eqn:g2_ext}
\end{align}
\end{widetext}
In order to evaluate the above expressions, one requires forms for $p_{\rm e}$, $\rho_{\rm e}(t)$ and $b(t)$.  Substituting the expressions for $M$ and $g_{2,{\rm e}}$ into the generalised Oppenheimer--Volkov equation (\ref{eq:OVgeneralised}), integrating and imposing the condition that the pressure is continuous across the boundary $b(t)$ and hence vanishes there, will yield an (integral) expression for $p_{\rm e}$ in terms of $\rho_{\rm e}(t)$, $H_{\rm e}(t)$, $b(t)$ and the (in general) time-dependent uniform fluid pressure $p_{\infty}(t)$ at large $r$ corresponding to the external cosmological model. One is free to specify $H_{\rm e}(t)$, $b(t)$ and $p_{\infty}(t)$, and the function $\rho_{\rm e}(t)$ may be determined from the following equations from NLH3 which remain valid in the exterior region:
\begin{align}
\rho_{\rm e}'(t) + 3H_{\rm e}(t) \left( \rho_{\rm e}(t) + p_{\infty}(t) \right) &= 0, \label{eq:cosmo1}\\
	H_{\rm e}^{2}(t)-\tfrac{8 }{3} \pi \rho_{\rm e}(t)- \tfrac{1}{3}\Lambda &= - \frac{k}{R^2(t)}. \label{eq:cosmo2}
\end{align}
We recognise (\ref{eq:cosmo1}) and (\ref{eq:cosmo2}) as the standard cosmological fluid evolution equation and the Friedmann equation, respectively. Moreover, as in NLH3, these can be combined in the usual manner to yield the dynamical cosmological field equation
\begin{equation}
H'_{\rm e}(t) + H_{\rm e}^2(t) -\tfrac{1}{3}\Lambda
= -\tfrac{4}{3}\pi(\rho_{\rm e}(t)+3p_\infty(t)),
\label{eq:cosmo3}
\end{equation}
which thus provides an expression for $\rho_{\rm e}(t)$ in terms of $H_{\rm e}(t)$ and $p_\infty(t)$. 

We are free to choose the boundary evolution $b(t)$, and it is most convenient to do this by defining the `vacuum Hubble parameter' $H_{\rm v}(t)$, such that
\begin{equation}
  g_{2,{\rm e}}(b(t),t)= b(t)H_{\rm v}(t).
  \label{eqn:Hv_defn}
\end{equation}
Equating the above expression with \eqref{eqn:g2_ext} evaluated on the boundary $b(t)$, one obtains
\begin{equation}
  \frac{b'(t)}{b(t)}= -\frac{H_{\rm v}(t) \rho_{\rm e}'(t)}{3\rho_{\rm
      e}(t) H_{\rm e}(t)}.
  \label{eqn:bdash}
\end{equation}
This then allows one to write \eqref{eqn:g2_ext} in the elegant form
\begin{equation}
  g_{2,{\rm e}}=rH_{\rm e}(t) - \frac{b^3(t)}{r^2} \left( H_{\rm
    e}(t)-H_{\rm v}(t) \right).
  %in notes 19/07/15
  \label{eqn:v_ext2}
\end{equation}
In a similar manner, one may write the expression \eqref{eqn:g1_ext} as
\begin{align}
    g^2_{1,{\rm e}} =1&-\frac{2m}{r} + (H_{\rm e}^2-\tfrac{8}{3}\pi
    \rho_{\rm e} -
    \tfrac{1}{3}\Lambda )r^2 \nonumber \\
&+ \frac{b^3}{r} (H_{\rm v} - H_{\rm e}) \left[2H_{\rm e} +
  \frac{b^3}{r^3} 
(H_{\rm v} - H_{\rm e}) \right],
  %in notes 19/07/15
    \label{eqn:g1_ext2}
\end{align}
where we have momentarily suppressed $t$-dependencies for brevity. It is worth noting that, for the special case of a pressureless fluid, one immediately has $p_{\rm e}=0$ and $f_{1,{\rm e}}=1$, so \eqref{eqn:bdash} becomes simply $b'(t)/b(t) = -H_{\rm v}(t)$.

\subsection{Vacuum region}

In the vacuum region, one has $\rho=p=0$ and so $L_rM=0$ and $L_tM=0$, which together imply $M$ is a constant, which we have denoted by $M_0$. As we found in our discussion of the Schwarzschild spacetime in Section~\ref{sec:tetradschwarzapp}, in a vacuum region the system of equations \eqref{eq:tetrels}--\eqref{eq:quantities} reduces just to the relationships \eqref{eq:tetrels} between the tetrad and spin-connection components and the definition of $M (= M_0)$ in \eqref{eq:quantities}, from which one finds that the quantity
\begin{equation}
  g_1^2- g_2^2 = 1-\frac{2 M_0}{r} - \tfrac{1}{3}\Lambda
  r^2 \equiv \alpha(r)
  \label{eqn:j_term}
\end{equation}
is a function of $r$ only. No further equations yield new information, and one thus has an under-determined system of equations that requires additional gauge-fixing to determine an explicit solution.

This occurs because in a vacuum region one clearly cannot choose the timelike unit Lorentz frame vector at each point to coincide with the fluid four-velocity at that point. As we did for the Schwarzschild spacetime, however, one may instead choose the timelike unit frame vector to coincide with the four-velocity $\mathbf{u}$ of some radially-moving test particle (which need not necessarily be in free-fall), so that its components in the tetrad frame are $[\hat{u}^a] = [1,0,0,0]$ and hence in the coordinate basis one has $[u^\mu] \equiv [\dot{t},\dot{r},\dot{\theta},\dot{\phi}]=[f_{1,{\rm v}},g_{2,{\rm v}},0,0]$, as previously. Moreover, this ensures that our previous physical interpretations of the tetrad and spin-connection components still hold.

Unlike in the Schwarzschild spacetime, however, there is no simplest or most natural choice for the class of radially-moving test particle to use. All one requires is that the boundary conditions discussed in Section~\ref{sec:bcs} hold at each boundary $a(t)$ and $b(t)$. It is most convenient to begin by choosing $g_{2,{\rm v}} = v(r,t)$, where $v$ may be an arbitrary function satisfying the boundary conditions $v(a(t),t)=g_{2,{\rm i}}(a(t),t)=a(t)H_{\rm i}(t)$ and $v(b(t),t)=g_{2,{\rm e}}(b(t),t)=b(t)H_{\rm v}(t)$. Then $g_{1,{\rm v}}$ is easily found from \eqref{eqn:j_term} and is also continuous at both boundaries. Finally, eliminating $G$ between the $L_rf_1$ and $L_t g_1$ equations in \eqref{eq:tetrels} gives the general relation
\begin{equation}
\partial_r(f_1g_1)+\frac{f_1^2}{g_2}\partial_tg_1=0,
\label{eqn:elimgrel}
\end{equation}
and using \eqref{eqn:j_term} to substitute for $g_1$, one then finds
\begin{equation}
\partial_r\left(f_{1,{\rm v}}\sqrt{\alpha(r)+v^2}\right) + f_{1,{\rm v}}^2 \frac{\partial_t
  v}{\sqrt{\alpha(r)+v^2}}=0,
\label{eqn:Lrf1_2}
\end{equation}
which may be straightforwardly solved for $f_{1,{\rm v}}$. Gathering these results together, in the vacuum region one thus has
\begin{equation}
  \begin{aligned}[b]
      g_{2,{\rm v}} &=v, \\
    g_{1,{\rm v}} &= \sqrt{\alpha(r)+v^2},\\
    f_{1,{\rm v}} &= \left( \sqrt{\alpha(r)+v^2} \int  \frac{\partial v /\partial t}{\left( \alpha(r)+v^2 \right)^{3/2}} \,dr  \right)^{-1}.
  \end{aligned}
  \label{eqn:fns_vac}
\end{equation}

Let is first consider the general case in which the fluid in the interior and exterior regions supports pressure. Suppose one specifies the form for $v(r,t_\ast)$ at some time $t_\ast$. One can then see from \eqref{eqn:Lrf1_2} that one also requires the profile $f_{1,{\rm v}}(r,t_\ast)$ in order to evolve $v$ in time. Thus, both $v(r,t_*)$ and $f_{1,{\rm v}}(r,t_*)$ need to be specified to determine the system. One should note, however, that there is no equation determining the time evolution of $f_{1,{\rm v}}$; hence $f_{1,{\rm v}}$ is free to take any value on any time slice, provided it satisfies the boundary conditions that both $f_1$ and $\partial_r f_1$ are continuous at each boundary, as shown in Section~\ref{sec:bcs}. Then the time evolution of $v$ is determined.

The situation is somewhat simpler for the case in which the fluid in the interior and exterior regions is pressureless, since one may take $f_1=1$ everywhere and at all times. Hence, if one specifies the form for $v(r,t_\ast)$ at some time $t_*$, one can use \eqref{eqn:Lrf1_2} to evolve $v$ in time \citep{Lasenby1999}. In this case, \eqref{eqn:Lrf1_2} becomes
\begin{equation}
\partial_tv+v\,\partial_rv = -\frac{M_0}{r}+\tfrac{1}{3}\Lambda r,
\end{equation}
which may also be derived directly by substituting the definition of $M$ in \eqref{eq:quantities} into $L_tM=0$.

In either case, with or without pressure, the exact choice of the initial profile $v(r,t_\ast)$ or $f_{1,{\rm v}}$ in the vacuum region has no physical effects on, for instance, the total redshift of a photon passing through the inhomogeneity \citep{Lasenby1999}, showing that the ambiguity in these functions is a gauge freedom and hence has no physical consequences.

\section{Birkhoff's theorem}
\label{sec:birkhoff}

One of our motivations for considering the generalised Swiss cheese model is to verify that Birkhoff's theorem holds in the vacuum region, despite having time-evolving interior and exterior regions. Although its usual interpretation has recently been brought into question for physical systems of this type \cite{Zhang2012}, we first demonstrate that Birkhoff's theorem does indeed hold using a traditional wholly metric-based approach, and then proceed to show that it also holds directly at the level of the tetrad components. In particular, we note that the considerations in this Section do not depend on the radial distribution of matter in the interior or exterior regions or its state of motion, provided spherical symmetry holds.

\subsection{Metric-based approach}

Our aim is to show that one may perform a transformation to new coordinates $\bar{t}$ and $\bar{r}$ that brings the line-element in the vacuum region into the standard Schwarzschild--de-Sitter form
\begin{equation}
ds^2 =
{\textstyle\left(1-\frac{2M_0}{\bar{r}}-\tfrac{1}{3}\Lambda\bar{r}^2\right)}
\,d\bar{t}^2 -
\frac{d\bar{r}^2}{\left(1-\frac{2M_0}{\bar{r}}-\tfrac{1}{3}\Lambda\bar{r}^2\right)}
- \bar{r}^2 d\Omega^2.
\label{eqn:sdsmetric}
\end{equation}

It is most instructive to begin by considering the general form \eqref{eq:generalmetric} of the line-element given in terms of the tetrad components. Since the angular part of \eqref{eq:generalmetric} already has same form as in \eqref{eqn:sdsmetric}, this suggests that one should consider a coordinate transformation of the form
\begin{equation}
t=t(\bar{t},\bar{r}),\qquad r=\bar{r},
\label{eqn:trans2}
\end{equation}
which is, in some sense, complementary to the coordinate transformation (\ref{eqn:transf}) considered previously. By analogy with our earlier discussion, however, although the radial coordinates coincide, we still label the new one as $\bar{r}$, since the partial derivatives $\partial/\partial r$ and $\partial/\partial\bar{r}$ will, in general, be different. 

It is straightforward to show that if the coordinate transformation \eqref{eqn:trans2} satisfies the conditions
\begin{equation}
\frac{\partial t}{\partial\bar{t}} = f_1g_1,\qquad
\frac{\partial t}{\partial\bar{r}} = -\frac{f_1g_2}{g_1^2-g_2^2},
\label{eqn:trans2derivs}
\end{equation}
then the line-element \eqref{eq:generalmetric} takes the diagonal form
\begin{equation}
ds^2 = (g_1^2-g_2^2)\,d\bar{t}^2 -
\frac{d\bar{r}^2}{g_1^2-g_2^2}-\bar{r}^2\,d\Omega^2.
\label{eqn:yametric}
\end{equation}
In order for such a transformation to be possible, however, one requires the derivatives \eqref{eqn:trans2derivs} to be consistent, i.e.\ they must satisfy $\partial^2 t/\partial\bar{t}\,\partial\bar{r} = \partial^2 t/\partial\bar{r}\,\partial\bar{t}$. Writing
\begin{align}
\frac{\partial}{\partial\bar{t}} &= \frac{\partial
  t}{\partial\bar{t}}\frac{\partial}{\partial t} +\frac{\partial
  r}{\partial\bar{t}}\frac{\partial}{\partial r}
= f_1g_1\frac{\partial}{\partial t}, \nonumber \\ 
\frac{\partial}{\partial\bar{r}} &=
\frac{\partial
  t}{\partial\bar{r}}\frac{\partial}{\partial t} +\frac{\partial
  r}{\partial\bar{r}}\frac{\partial}{\partial r}
=-\frac{f_1g_2}{g_1^2-g_2^2}\frac{\partial}{\partial t} +
\frac{\partial}{\partial r},
\end{align}
and using the general relation \eqref{eqn:elimgrel}, one finds that the derivatives are consistent only in the case where $g_1^2-g_2^2$ is a function of $r$ alone. As shown in \eqref{eqn:j_term}, this requirement is satisfied in the vacuum region and, moreover, the resulting line-element \eqref{eqn:yametric} then indeed takes the Schwarzschild--de-Sitter form \eqref{eqn:sdsmetric}, hence verifying Birkhoff's theorem.

\subsection{Tetrad-based approach}

When working at the level of the metric, one is insensitive to rotations of the local Lorentz frames. Indeed, although the coordinate transformation \eqref{eqn:trans2}--\eqref{eqn:trans2derivs} brings the line-element into the Schwarzschild-de-Sitter form \eqref{eqn:sdsmetric} in the vacuum region, the corresponding tetrad components are {\em not} of the standard `diagonal' form \eqref{eqn:schwarztetrad} (together with the Newtonian gauge condition $f_2=0$ and after the inclusion of obvious additional terms resulting from a non-zero cosmological constant $\Lambda$).

Indeed, since the tetrad components transform as ${\bar{e}_a}^{\phantom{a}\mu} = (\partial\bar{x}^\mu/\partial x^\nu) {e_a}^\nu$ under a general coordinate transformation, it is straightforward to show that
\begin{equation}
\bar{f}_1 = \frac{g_1}{g_1^2-g_2^2},\quad
\bar{f}_2 = \frac{g_2}{g_1^2-g_2^2},\quad
\bar{g}_1 = g_1,\quad
\bar{g}_2 = g_2,
\end{equation}
so that the new tetrad components do not even satisfy the Newtonian gauge condition $\bar{f}_2=0$. Nonetheless, the tetrad components can be transformed into the standard diagonal Schwarzschild form \eqref{eqn:schwarztetrad} by performing a (temporally- and radially-dependent) rotation of the local Lorentz frames, which automatically leaves the metric unchanged. Under such a transformation, the tetrad components become ${\bar{\bar{e}}_a}^{\mu} = {\Lambda_a}^b {\bar{e}_b}^{\phantom{b}\mu}$, where 
\begin{equation}
\left[{\Lambda^a}_b\right] = 
\left(
\begin{array}{cccc}
\phantom{-}\cosh\psi & -\sinh\psi & 0 & 0 \\
-\sinh\psi & \phantom{-}\cosh\psi & 0 & 0 \\
0 & 0 & 1 & 0 \\
0 & 0 & 0 & 1
\end{array}
\right)
\end{equation}
and $\psi$ denotes the rapidity of the radial boost at each event. Thus, one finds
\begin{align}
\bar{\bar{f}}_1 & = (g_1\cosh\psi + g_2\sinh\psi)/(g_1^2-g_2^2),
\nonumber \\
\bar{\bar{f}}_2 & = (g_1\sinh\psi + g_2\cosh\psi)/(g_1^2-g_2^2),
\nonumber \\
\bar{\bar{g}}_1 & = g_1\cosh\psi + g_2\sinh\psi, \nonumber \\
\bar{\bar{g}}_2 & = g_1\sinh\psi + g_2\cosh\psi.
\end{align}
Hence, by choosing the rapidity $\psi$ at each event such that $g_1\sinh\psi + g_2\cosh\psi=0$, one immediately ensures that $\bar{\bar{f}}_2=\bar{\bar{g}}_2=0$. Since $\cosh^2\psi-\sinh^2\psi=1$, the required rapidity is given by $\sinh\psi = \pm g_2/\sqrt{g_1^2-g_2^2}$. Therefore, one further obtains
\begin{equation}
\bar{\bar{f}}_1 = \frac{1}{\sqrt{g_1^2-g_2^2}},\qquad
\bar{\bar{g}}_1 = \sqrt{g_1^2 - g_2^2},
\end{equation}
where $g_1^2 - g_2^2$ is given by (\ref{eqn:j_term}), and thus we recover the standard diagonal form of the tetrad for the Schwarzschild solution. Finally, it is worth noting that the speed of the boost in the radial direction is given at each point by $\tanh\psi = -g_2/g_1$.

\subsection{Comparison with previous work}
\label{subsec:comparison}

The validity of Birkhoff's theorem, or at least its usual interpretation, has recently been brought into question by \cite{Zhang2012}.  To be clear, Birkhoff's theorem states that there exist coordinates for which the metric in a vacuum region surrounding any spherically-symmetric matter distribution takes the standard Schwarzschild (de-Sitter) form with parameter $M_0$ equal to the enclosed interior mass, even when the vacuum region is itself embedded in an exterior spherically-symmetric matter distribution. Indeed, we have just verified this in the previous Section, and shown further that the theorem also holds directly at the level of the tetrad components. An immediate corollary of Birkhoff's theorem is that, if the enclosed mass $M_0=0$, there exist coordinates for which the metric in the vacuum region takes the standard Minkowski (de-Sitter) form.

As pointed out by \cite{Zhang2012}, however, a common misinterpretation of Birkhoff's theorem is that the gravitational field anywhere inside a spherically-symmetric matter distribution is determined only by the enclosed mass. While this is true in Newtonian gravity, it does not hold in general relativity. This point is illustrated in \cite{Zhang2012} by considering the metric corresponding to a static thin spherical shell of mass $m_{\rm s}$ and coordinate radius $r=r_{\rm s}$ surrounding a spherical central object of mass $m_{\rm i}$ centered on the origin. Since some aspects of the original calculation are unclear, we briefly re-examine this scenario here, in which the cosmological constant is assumed to vanish.

Following \cite{Zhang2012}, we begin by assuming a static, diagonal line-element of the form
\begin{equation}
ds^2 = A(r)\,dt^2 - B(r)\,dr^2 - r^2\,d\Omega^2,
\end{equation}
where the functions $A(r)$ and $B(r)$ are arbitrary (note that our definitions of $A$ and $B$ are swapped relative to the line-element used in \cite{Zhang2012}). The thin spherical shell is modelled by the artifice of first considering a shell of finite thickness with mass density $\rho(r)$, supporting tangential pressure $p(r)$ but no radial pressure component. Although not discussed by \cite{Zhang2012}, it is only by assuming the shell is comprised of such a peculiar `fluid', with zero radial stress but non-zero compressive stress in the tangential directions, that a static system is possible. In particular, if the radial stress did not vanish, then it would be necessary to match it to zero at the boundaries with the interior and exterior vacua. 

Nonetheless, proceeding with the above model, the components of the energy-momentum tensor within the shell are $[T_{\mu\nu}] = \mbox{diag}(\rho A,0,p r^2,p r^2\sin^2\theta)$, and the equations of motion yield
\begin{align}
\frac{dA}{dr} &= \frac{2AM}{r(r-2M)}, \label{eqn:dadr}\\
B&=\left(1-\frac{2M}{r}\right)^{-1}, \label{eqn:bequals}\\
p & = \frac{M\rho}{2(r-2M)}, \label{eqn:pequals}
\end{align}
where $M(r) = \int_0^r 4\pi \bar{r}^2\rho(\bar{r})\,d\bar{r}$ is the total mass enclosed within coordinate radius $r$. In particular, in vacuum regions, where $\rho(r)=0$ and $M(r) = \mbox{constant}$, the equation \eqref{eqn:dadr} may be integrated immediately to give
\begin{equation}
A(r) = C\left(1-\frac{2M}{r}\right),
\end{equation}
where $C$ is a constant. Thus, coupled with the expression \eqref{eqn:bequals} for $B$, one sees that the metric in vacuum regions is of Schwarzschild form, but with a time coordinate rescaled by a constant factor $\sqrt{C}$ as compared to the proper time of a stationary observer at infinity.

The constant $C$ must be determined by applying the appropriate boundary conditions in the presence of the thin spherical shell lying between the origin and infinity.  In this case, $\rho(r)$ has a $\delta$-function at $r=r_{\rm s}$ (corresponding to an infinitesimally thin shell). Thus, at this radius, $M$ has a step and equations \eqref{eqn:dadr}--\eqref{eqn:pequals} show that $p$ has a $\delta$-function and $B$ and $dA/dr$ have a step. The last condition means that $A$ must be continuous at $r=r_{\rm s}$. Thus, since we require $A \to 1$ at infinity, the appropriate constant $C$ to use in the vacuum region interior to the shell must satisfy
\begin{equation}
C\left(1-\frac{2m_{\rm i}}{r_{\rm s}}\right)=1-\frac{2(m_{\rm i} +
  m_{\rm s})}{r_{\rm s}},
\end{equation}
and so, in this region, the line-element is 
\begin{equation}
ds^2 = 
{\textstyle\frac{r_{\rm s}-2(m_{\rm i} +m_{\rm s})}{r_{\rm s} - 2m_{\rm
    i}}\left(1-\frac{2m_{\rm i}}{r}\right)}\,dt^2-\frac{dr^2}{\left(1-\frac{2m_{\rm i}}{r}\right)}-r^2\,d\Omega^2.
%A(r) = \frac{r_{\rm s}-2(m_{\rm i} +m_{\rm s})}{r_{\rm s} - 2m_{\rm
%    i}}\left(1-\frac{2m_{\rm i}}{r}\right).
\end{equation}

The form of the line-element, in particular the rate at which clocks run, in the vacuum region interior to the shell thus depends on both the mass $m_{\rm s}$ and location $r_{\rm s}$ of the shell, i.e.\ the gravitational field in this region depends on the matter distribution external to it. This is unsurprising, of course, since the presence of the shell puts the interior vacuum region into a deeper potential well with respect to infinity than would be the case without the shell. Thus, one would expect clocks to run more slowly in this region with the shell place than without it.  The same is clearly true if one considers a hollow cavity by setting $m_{\rm i}=0$, although the spacetime inside the cavity is then Minkowski rather than Schwarzschild. Clearly, these considerations do not contradict Birkhoff's theorem or its corollary, however, since one can perform a simple constant rescaling of the time coordinate to recover the Schwarzschild line-element. Nonetheless, as pointed out by \cite{Zhang2012}, it would be incorrect to perform such a rescaling, or equivalently set $C=1$ in the interior vacuum region, while leaving the line-element outside the shell unchanged, since this would lead to an unphysical discontinuity the time coordinate, and hence $A(r)$, across the shell.  It is unclear, however, whether any of the works criticised in \cite{Zhang2012} ever advocate carrying out this erroneous procedure.

A further example of the gravitational field at some radius in a spherically-symmetric matter distribution being determined by material external to that radius is provided by the system analysed in \cite{NLH3}. As mentioned previously, this system corresponds to setting $b(t)=a(t)$ in the generalised Swiss cheese model discussed in Section~\ref{sec:gscm}, so that there is no vacuum region. In \cite{NLH3}, this system is analysed separately using Newtonian gravity and general relativity. The former case, the Newtonian gravitational potential in the interior region is found to be independent of the properties of the exterior region, whereas the general-relativistic calculation shows that some of the tetrad components, and hence metric elements, in the interior do depend on the properties of the exterior region, such as its density $\rho_{\rm e}(t)$ and Hubble parameter $H_{\rm e}(t)$.

\section{$R_{\rm h} = ct$ cosmology}
\label{sec:rctmodel}

The current standard model of cosmology is based on the $\Lambda$CDM model, which provides a good fit to a wide range of cosmological observations. As pointed out by \cite{Melia2003}, however, for the best-fit $\Lambda$CDM model, the present-day Hubble distance is broadly consistent with $ct_0$ to within observational uncertainties, where $t_0$ is the current cosmic epoch. This suggests that the universe has expanded by an amount similar to what would have occurred had the expansion rate been constant.  In the $\Lambda$CDM model, this correspondence is a peculiar coincidence, particularly since this situation should occur only once in the history of the universe; that we observe it to hold at the present epoch is thus intriguing.

It was therefore proposed by \cite{Melia2007,Melia2009,MeliaShevchuk2012} that this correspondence is not coincidental, but should be satisfied at all cosmic times $t$.  The resulting cosmological model, known as the `$R_{\rm h}=ct$' model, has received considerable attention over the last few years, since it has been claimed to be favoured over the standard $\Lambda$CDM (and its variant $w$CDM with $w\neq -1$) by most observational data \citep{MeliaMaier2013,Wei2013,Wei2014a,Wei2014b,Wei2015,Melia_etal2015}.

These claims have, however, recently been brought into question.  For example, the $R_{\rm h}=ct$ model requires the deceleration parameter $q(z)=0$ at all redshifts, but it is pointed out in \cite{Bilicki2012} that recent supernovae and baryon acoustic oscillations observations rule out $q_0=0$ at high significance. Moreover, the most recent observations of CMB anisotropies by the {\em Planck} mission suggest that the primary motivation for the $R_{\rm h}=ct$ model, namely that the age of the universe is $1/H_0$, can be ruled out at greater than the 99 per cent confidence level, with the data favouring $R_{\rm h} = \left(1.05 \pm 0.02\right)ct$ at the current epoch \citep{vanOirschot2014}. Finally, as we show below, in the spatially-flat $\Lambda=0$ FRW spacetime assumed in the $R_{\rm h}=ct$ model, the comoving Hubble radius is constant in time. This contradicts the basic picture of perturbation generation and evolution according to inflationary cosmological models, in which the comoving Hubble radius decreases during inflation and then increases again afterwards. This enables the generation of `superhorizon' modes that later re-enter the Hubble radius coherently as it grows. This leads in turn to the characteristic acoustic peak structure observed in the power spectrum of CMB anisotropies and to baryon acoustic oscillations; these well-established phenomena would be very difficult to explain if the comoving Hubble radius were constant.

In addition to observational objections, the validity of the physical argument underlying the $R_{\rm h}=ct$ model has also been criticised by a number of authors \citep{vanOirschot2010,LewisvanOirschot2012,Lewis2013,Mitra2014}.  These and other criticisms are claimed to have been addressed by \cite{Bikwa2012, Melia2012c} (see also \cite{Melia2015} and references therein), but these responses often merely restate the original claims for the theory and do not address many of the specific concerns raised.  Having examined spherically-symmetric solutions in general relativity in previous sections, including cosmological solutions, we take the opportunity here to re-evaluate the theoretical arguments originally used to arrive at the $R_{\rm h}=ct$ model.

In \cite{Melia2007,Melia2009,MeliaShevchuk2012}, one is first invited to imagine precisely the system illustrated in Figure~\ref{fig:model_diag}, namely the generalised Swiss cheese model, with an observer located at the origin and assuming $\Lambda=0$. By appealing to Birkhoff's theorem and its corollary, it is then pointed out correctly that the metric in the vacuum region may be written in the standard Schwarzschild form. It is further claimed, however, that the matter distribution in the exterior region therefore has no dynamical influence on the region contained within it. As outlined in Section~\ref{subsec:comparison}, this does not necessarily hold in general relativity.

Nonetheless, this issue is not relevant to Melia's subsequent argument, since it later transpires that his appeal to Birkhoff's theorem is merely to justify that, in a standard homogeneous and isotropic cosmological model, one may determine the motion of a fluid particle at some coordinate radius $r$ relative to an arbitrary origin by ignoring the gravitational effect of the fluid lying outside that radius, which does hold in this particular case. Thus, the scenario Melia considers is the standard homogeneous and isotropic cosmological model, and not the generalised Swiss cheese model illustrated in Figure~\ref{fig:model_diag} that we are first invited to imagine. Indeed, it is clear that the latter is not a viable cosmological model in general. Aside from being inhomogeneous overall, the requirement that the pressure be continuous across the boundaries means that, even if one assumes that each non-vacuum region is homogeneous (i.e.\ with uniform density and pressure), they cannot be matched on the vacuum region unless the pressure vanishes everywhere; thus one could consider only dust models of this type.

Melia's main proposal begins with the introduction of the `gravitational radius' $R_{\rm h}(t)$, which is defined by the requirement that
\begin{equation}
\frac{2M(R_{\rm h}(t),t)}{R_{\rm h}(t)} \equiv 1,
\label{eqn:rhdef}
\end{equation}
where, as previously, we adopt natural units $c=G=1$ and $M(r,t)=\frac{4}{3}\pi\rho(t)r^3$ is the mass-energy contained with a sphere of physical radius $r$ at time $t$. Indeed, substituting this form for $M(r,t)$ into \eqref{eqn:rhdef}, one immediately finds that $1/R^2_{\rm h}(t) = \frac{8}{3}\pi\rho(t)$. The key point in Melia's argument is the assertion that, in order to satisfy Weyl's postulate, the comoving gravitational radius $\hat{r}_{\rm h}(t)\equiv R_{\rm H}(t)/S(t)$ should be {\em independent}\/ of $t$, where $S(t)$ is the scale factor defined in Section~\ref{sec:tetradFRWapp}.  In other words, each fluid particle has  a {\em fixed} value of $\hat{r}_{\rm h}$. 

\begin{table*}[t!]
 \caption{Evolution of the comoving radius $a(t)$, its velocity
   $\dot{a}(t)$, the mass-energy $M(t)$ enclosed within $a(t)$, the
   Schwarzschild radius (or radii) $R_{\rm S}(t)$ defined in
   \eqref{eqn:rsdef} and the Hubble radius $R_{\rm H}(t)\equiv 1/H(t)$
   for a selection of analytical spatially-flat $(k=0)$ expanding
   cosmological models, with cosmological constant $\Lambda$ and fluid
   equation-of-state parameter $w$, whose evolution is determined by
   the Hubble parameter $H(t)$. The values of $w$ considered
   correspond to dust $(w=0)$, radiation $(w=\tfrac{1}{3})$ and
   Melia's zero-active-mass condition $(w=-\tfrac{1}{3})$. Note that,
   for $\Lambda\neq 0$, the two positive solutions for $R_{\rm S}(t)$
   are valid only if $1-9M^2(t)\sqrt{\Lambda} > 0$; otherwise there is
   no positive solution.
\label{table:horizons}}
 \begin{ruledtabular}
 \begin{tabular}{c|ccc|cc}
& \multicolumn{3}{c|}{$\Lambda=0$} &  \multicolumn{2}{c}{$\Lambda\neq 0$} \\ [1mm]
& $w=0$ & $w=\tfrac{1}{3}$ & $w=-\tfrac{1}{3}$ \qquad & $w=0$ &
   $w=\tfrac{1}{3}$ \\ [1mm]
\hline \\[-2mm]
$H(t)$ &$\frac{2}{3}t^{-1}$ & $\frac{1}{2}t^{-1}$ & $t^{-1}$ & 
$\sqrt{\frac{\Lambda}{3}}\coth\left(\frac{\sqrt{3\Lambda}}{2}t\right)$
&
$\sqrt{\frac{\Lambda}{3}}\coth\left(\frac{2\sqrt{\Lambda}}{3}t\right)$
\\[2mm]
$a(t)$ & $2M_0^{1/3}t^{2/3}$ & $2M_0^{1/3}t^{1/2}$ &
$2M_0^{1/3}t$ & $2M_0^{1/3}\sinh^{2/3}\left(\frac{\sqrt{3\Lambda}}{2}t\right)$
& $2M_0^{1/3}\sinh^{1/2}\left(\frac{2\sqrt{\Lambda}}{3}t\right)$ \\[2mm]
$\dot{a}(t)$ & $\tfrac{4}{3}M_0^{1/3}t^{-1/3}$ & $M_0^{1/3}t^{-1/2}$ &
%$2M_0^{1/3}$ & $2\sqrt{\frac{\Lambda}{3}}M_0^{1/3}\coth\left(\frac{\sqrt{3\Lambda}}{2}t\right)\sinh^{2/3}\left(\frac{\sqrt{3\Lambda}}{2}t\right)$
$2M_0^{1/3}$ & $2\sqrt{\frac{\Lambda}{3}}M_0^{1/3}\frac{\cosh\left(\frac{\sqrt{3\Lambda}}{2}t\right)}{\sinh^{1/3}\left(\frac{\sqrt{3\Lambda}}{2}t\right)}$
 &
%$2\sqrt{\frac{\Lambda}{3}}M_0^{1/3}\coth\left(\frac{2\sqrt{\Lambda}}{3}t\right)\sinh^{1/2}\left(\frac{2\sqrt{\Lambda}}{3}t\right)$
$2\sqrt{\frac{\Lambda}{3}}M_0^{1/3}
\frac{\cosh\left(\frac{2\sqrt{\Lambda}}{3}t\right)}{\sinh^{1/2}\left(\frac{2\sqrt{\Lambda}}{3}t\right)}$
\\[2mm]
$M(t)$ & $\frac{16}{9}M_0$ & $M_0 t^{-1/2}$ & $4M_0t$ &
$\tfrac{4}{3}M_0\Lambda$ &
$\frac{4M_0\Lambda}{3\sinh^{1/2}\left(\frac{2\sqrt{\Lambda}}{3}t\right)}$\\[2mm]
$R_{\rm S}(t)$ & $\frac{32}{9}M_0$ & $2M_0t^{-1/2}$ & $8M_0t$
& 
$\frac{2}{\sqrt{\Lambda}}\cos\left[\tfrac{\pi}{3}\pm\frac{1}{3}
\tan^{-1}
\left( -\frac{\sqrt{1-9M^2(t)\Lambda}}{3M(t)\sqrt{\Lambda}}\right)\right]$
&  
$\frac{2}{\sqrt{\Lambda}}\cos\left[\tfrac{\pi}{3}\pm\frac{1}{3}
\tan^{-1}
\left( -\frac{\sqrt{1-9M^2(t)\Lambda}}{3M(t)\sqrt{\Lambda}}\right)\right]$
\\[2mm]
$R_{\rm H}(t)$ & $\frac{3}{2}t$ & $2t$ & $t$ & $1/H(t)$ & $1/H(t)$ \\
\end{tabular}
\end{ruledtabular}
\end{table*}

Then, following Melia's assumption that $k=0=\Lambda$, the standard Friedmann equation \eqref{eq:tetradcosmo2} immediately allows one to make the identification $R_{\rm h}(t) = 1/H(t)$, which is the Hubble radius. Since $H(t)=\partial_t S(t)/S(t)$, the comoving gravitational (Hubble) radius in this case is given by
\begin{equation}
\hat{r}_{\rm h}(t) = \frac{1}{H(t)S(t)} = \frac{1}{\partial_t S(t)}.
\end{equation}
The requirement that $\hat{r}_{\rm h}(t)$ is constant therefore implies $S(t) \propto t$ and so $R_{\rm h}(t)= 1/H(t) = t$ or, on momentarily abandoning natural units, one obtains the eponymous $R_{\rm h}(t)=ct$. It is worth noting that substituting $H(t)=1/t$ into the dynamical cosmological field equation \eqref{eq:FRWaccel} yields Melia's so-called `zero active mass' condition $\rho+3p = 0$, which is equivalent to demanding that the overall cosmological fluid has the equation-of-state parameter $w=-\tfrac{1}{3}$ at all epochs.

The argument given in \cite{MeliaShevchuk2012} for supposing that $\hat{r}_{\rm H}(t)=\mbox{constant}$ is that, according to Weyl's postulate, any proper distance in an FRW spacetime must be the product of the scale factor $S(t)$ and some fixed co-moving radial coordinate, and that the definition of $R_{\rm h}(t)$ as a gravitational radius in \eqref{eqn:rhdef} implies that it must be a proper distance. No real justification is given for this latter assertion. Indeed, we show in Section~\ref{sec:horizons} below that the definition \eqref{eqn:rhdef} does {\em not}, in fact, imply that $R_{\rm h}(t)$ is a proper distance and hence one does {\em not} require $\hat{r}_{\rm h}(t)=\mbox{constant}$.  Nonetheless, in trying to understand the motivation for Melia's original assertion, it is worth noting that such a conclusion might be reached from the following {\em erroneous} line of reasoning, which, although not articulated as such in \cite{MeliaShevchuk2012}, has some resonances with the discussion given there.  

In a homogeneous and isotropic model, consider a fluid particle located at a distance $R_{\rm h}(t_\ast)$ from an arbitrary origin at some cosmic time $t=t_\ast$.  The definition (\ref{eqn:rhdef}) implies that $R_{\rm h}(t_\ast)$ is equal to the Schwarzschild radius of the mass of fluid contained with the radius $R_{\rm h}(t_\ast)$. Invoking (the corollary of) Birkhoff's theorem, the fluid external to this radius has no gravitational effect on the motion of the particle.  Thus, as the universe evolves, one might suspect that it is impossible for the fluid particle to move to a radius larger than $R_{\rm h}(t_\ast)$, since to do so would require it to escape the Schwarzschild horizon. Conversely, if the particle were to move to a radius smaller than $R_{\rm h}(t_\ast)$, then the original mass of fluid would be wholly contained within its Schwarzschild radius. Invoking (the corollary of) Birkhoff's theorem once again, one might suspect that the fluid would therefore inevitably undergo gravitational collapse. To avoid these two scenarios, one could thus be led to suppose that $R_{\rm h}(t) \propto S(t)$ and hence $\hat{r}_{\rm h}(t)=\mbox{constant}$.

In the next Section, we show that the requirement $\hat{r}_{\rm h}(t)=\mbox{constant}$ does {\em not} follow from the definition~\eqref{eqn:rhdef} and that the two scenarios outlined above are not a cause for concern. In doing so, our investigations below also elucidate the behaviour of a number of `horizons' during the general-relativistic evolution of a spherically-symmetric self-gravitating matter distribution; although straightforward and interesting, this behaviour does not appear to have been widely discussed in the literature.

Before moving on, however, we note that Melia has recently presented a new and more explicit theoretical argument for the $R_{\rm h}=ct$ model \cite{Melia2016}, in which he claims that its central feature of requiring zero active mass $\rho+3p=0$ at all epochs is a consequence of the symmetries of the FRW spacetime. In particular, it is claimed that assuming the general, spherically-symmetric (but radially varying) metric, solving the Einstein field equations, and then imposing homogeneity and isotropy yields an extra condition, namely vanishing active mass, which is lost if one adopts the usual procedure of first imposing the conditions of homogeneity and isotropy on the metric and then solving the Einstein equations. This claim is shown to be false in \cite{Kim2016}, where it is demonstrated that it originates from an obvious flaw in the logic of Melia's argument. This is contested by Melia in \cite{Melia2016a}, although this response largely just restates the original argument in \cite{Melia2016} and does not address the specific concerns raised in \cite{Kim2016}. We simply note here that in Sections~\ref{sec:tetradFRWapp} and~\ref{sec:LTBFRWapp}, we have carried out the process of imposing homogeneity and isotropy on the solutions of the Einstein field equations for a general, spherically-symmetric system, obtained using our tetrad-based approach and the LTB model, respectively. In both cases, we arrive at the standard FRW metric and the usual cosmological field equations, without encountering any requirement for vanishing active mass.

\section{Evolution of horizons}
\label{sec:horizons}

In a homogeneous and isotropic cosmological model, let us consider an imaginary spherical boundary of radius $a(t)$ that is comoving with the fluid and centred on some arbitrary origin. From the discussion in Section~\ref{sec:gscm}, the equation of motion for $a(t)$ is $L_ta(t) = \dot{a} = H(t)a(t)$ (since $f_1=1$ in this case), where $H(t)$ is the Hubble parameter characterising the evolution of the fluid. If $M(t)$ denotes the mass-energy contained within this sphere, we define the corresponding Schwarzschild radius $R_{\rm S}(t)$ as the solution of
\begin{equation}
1-\frac{2M(t)}{R_{\rm S}(t)} -\tfrac{1}{3}\Lambda R^2_{\rm S}(t) \equiv 0,
\label{eqn:rsdef}
\end{equation}
which clearly reduces to $R_{\rm S}(t) \equiv 2M(t)$ if $\Lambda=0$. We also define the Hubble radius $R_{\rm H}(t) \equiv 1/H(t)$; note that this coincides with Melia's gravitational radius $R_{\rm h}(t)$ in the case $k=0=\Lambda$, but differs from it in more general cosmological models (although such models were not considered by Melia). 

We now consider the behaviour of $a(t)$, $R_{\rm S}(t)$ and $R_{\rm H}(t)$ for a selection of analytical spatially-flat $(k=0)$ expanding cosmological models, with cosmological constant $\Lambda$ and fluid equation-of-state parameter $w$, whose evolution is determined by the Hubble parameter $H(t)$.  The results are presented in Table~\ref{table:horizons}.  The values of $w$ considered correspond to dust $(w=0)$, radiation $(w=\tfrac{1}{3})$ and Melia's zero-active-mass condition $(w=-\tfrac{1}{3})$. It is worth noting that it is only in the case $w=0$ that the mass-energy $M(t)$ contained within the comoving radius $a(t)$ is constant; for other values of $w$ the presence of non-zero pressure means that the fluid does work and hence $M(t)$ changes with time.

\begin{figure}
\includegraphics[width=\linewidth]{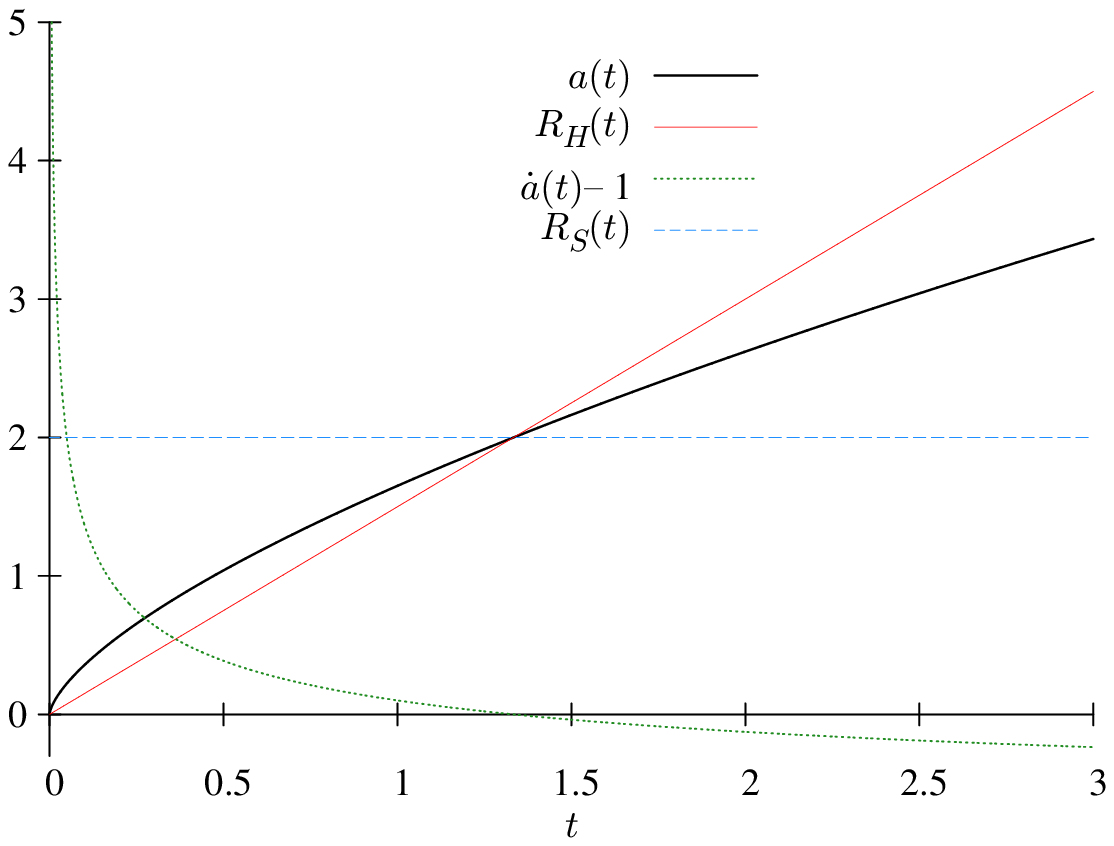}\\[1.65cm]
\includegraphics[width=\linewidth]{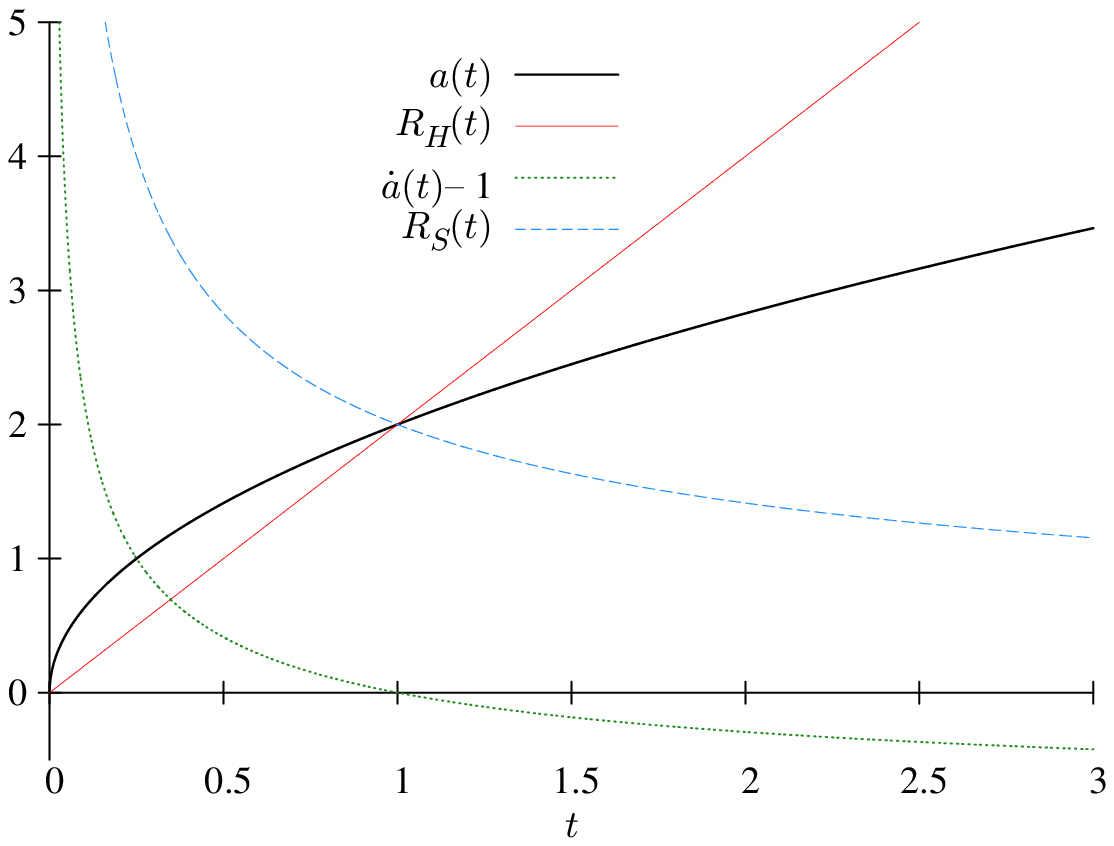}\\[1.65cm]
\includegraphics[width=\linewidth]{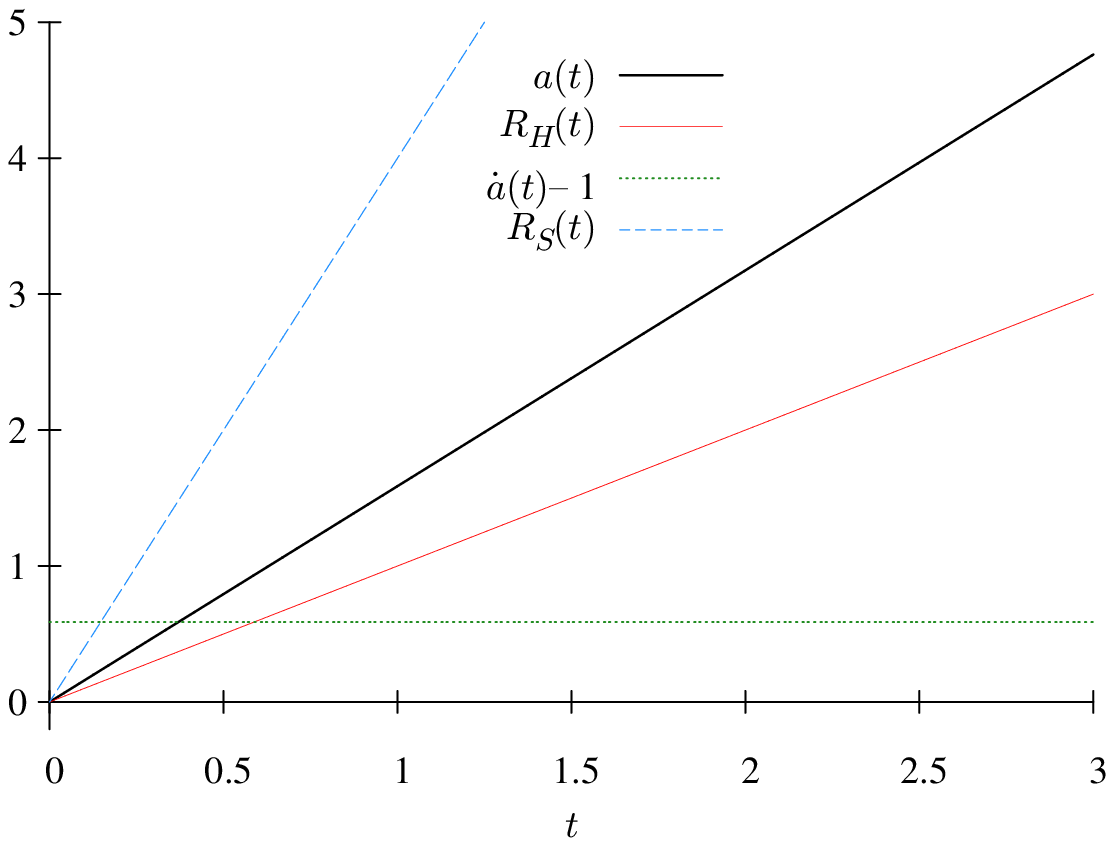}
\caption{Behaviour of quantities listed in Table~\ref{table:horizons}
  for the case $\Lambda=0$. Top: dust ($w=0$ with
  $M_0=\tfrac{9}{16}$), middle: radiation ($w=\tfrac{1}{3}$ with $M_0
  = 1$), bottom: zero active mass ($w=-\tfrac{1}{3}$ with
  $M_0=\tfrac{1}{2}$).\label{fig:horizons1}}
\end{figure}

For $\Lambda=0$, the behaviour of the quantities listed in the table is shown in Figure~\ref{fig:horizons1}.
For both the dust and radiation cases, one sees that $a(t)$, $R_{\rm S}(t)$ and $R_{\rm H}(t)$ cross at a single point that we denote by $t=t_\ast$, which coincides with the velocity of the fluid dropping to $\dot{a}(t)=1$ (or $c$ in standard units). This behaviour is quite general and holds for any positive value of the parameter $M_0$ in Table~\ref{table:horizons}. Thus, the comoving radius $a(t)$ initially lies outside the Hubble radius. Indeed, one may consider the superluminal fluid velocity during this initial period as being allowed by the `superhorizon' (or, more correctly, super-Hubble-radius) nature of $a(t)$. In any case, $a(t)$ enters the Hubble radius $R_{\rm H}(t)$ at precisely the same moment as it exits the Schwarzschild radius $R_{\rm S}(t)$; it is allowed to do the latter, since the fluid at the boundary is moving at speed $c$ at this instant.  Thus, one has two `horizon crossings' taking place simultaneously and in opposite directions, which is not usually pointed out in the literature. Moreover, one sees that there is no reason for the Hubble radius to be comoving, which contradicts the central assumption of the $R_{\rm h}=ct$ model. It is also worth noting that in the case of dust, for which the pressure vanishes everywhere, one can consistently `cut-off' the fluid at the boundary $a(t)$, and thereby consider a finite fluid ball surrounded by vacuum; the results for this case are precisely those given above.

The bottom panel in Figure~\ref{fig:horizons1} shows the behaviour for the case $w=-\frac{1}{3}$, which corresponds to the zero active mass condition required by the $R_{\rm h}=ct$ model. Clearly, in this case, the radii $a(t)$, $R_{\rm S}(t)$ and $R_{\rm H}(t)$ all depend linearly on $t$. If one chooses $M_0=\frac{1}{8}$, then one obtains the special case in which $a(t)=R_{\rm S}(t)=R_{\rm H}(t)$ and $\dot{a}(t)=1$ at all times. When $M_0 > \tfrac{1}{8}$, one has $\dot{a}(t)>1$ and $a(t)$ lies inside the Schwarzschild radius and outside the Hubble radius at all times (this is illustrated in Figure~\ref{fig:horizons1}, for which $M_0=\tfrac{1}{2}$). Conversely, if $M_0 < \tfrac{1}{8}$,  one has $\dot{a}(t)<1$ and $a(t)$ lies outside the Schwarzschild radius and in side the Hubble radius for all $t$.

\begin{figure}
\includegraphics[width=\linewidth]{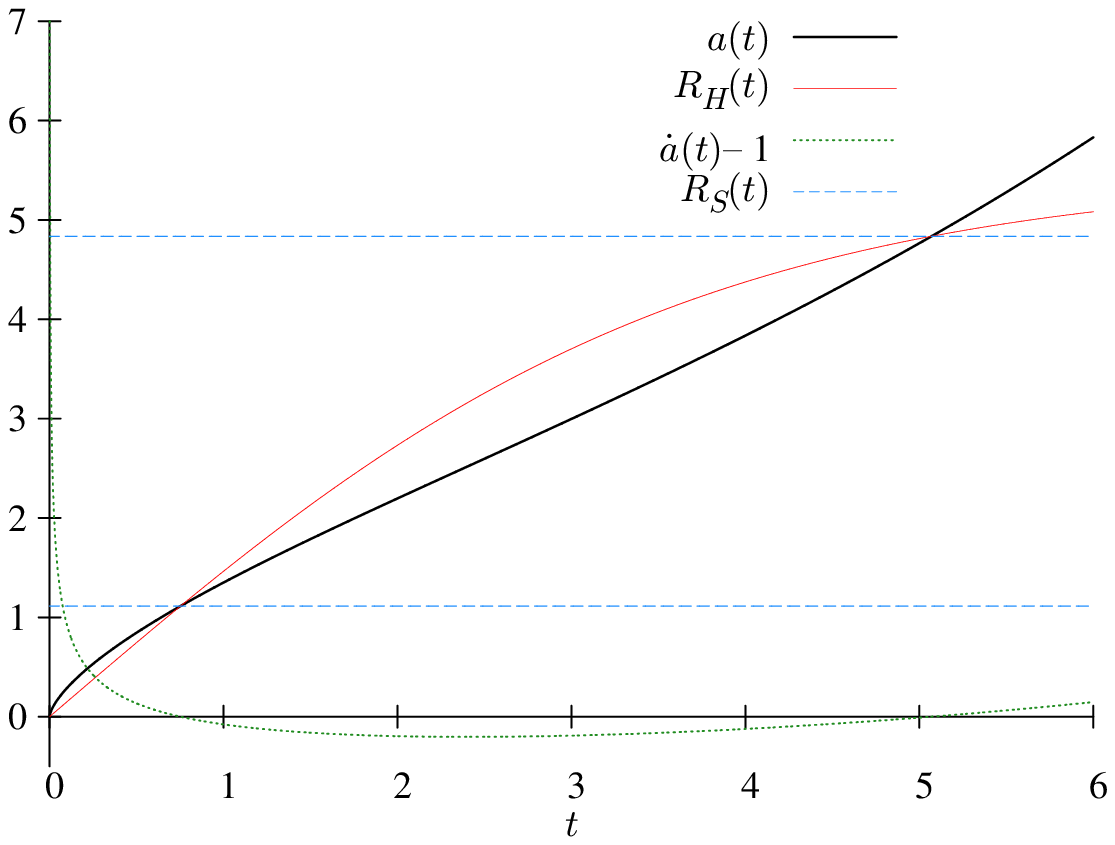}\\[1.65cm]
\includegraphics[width=\linewidth]{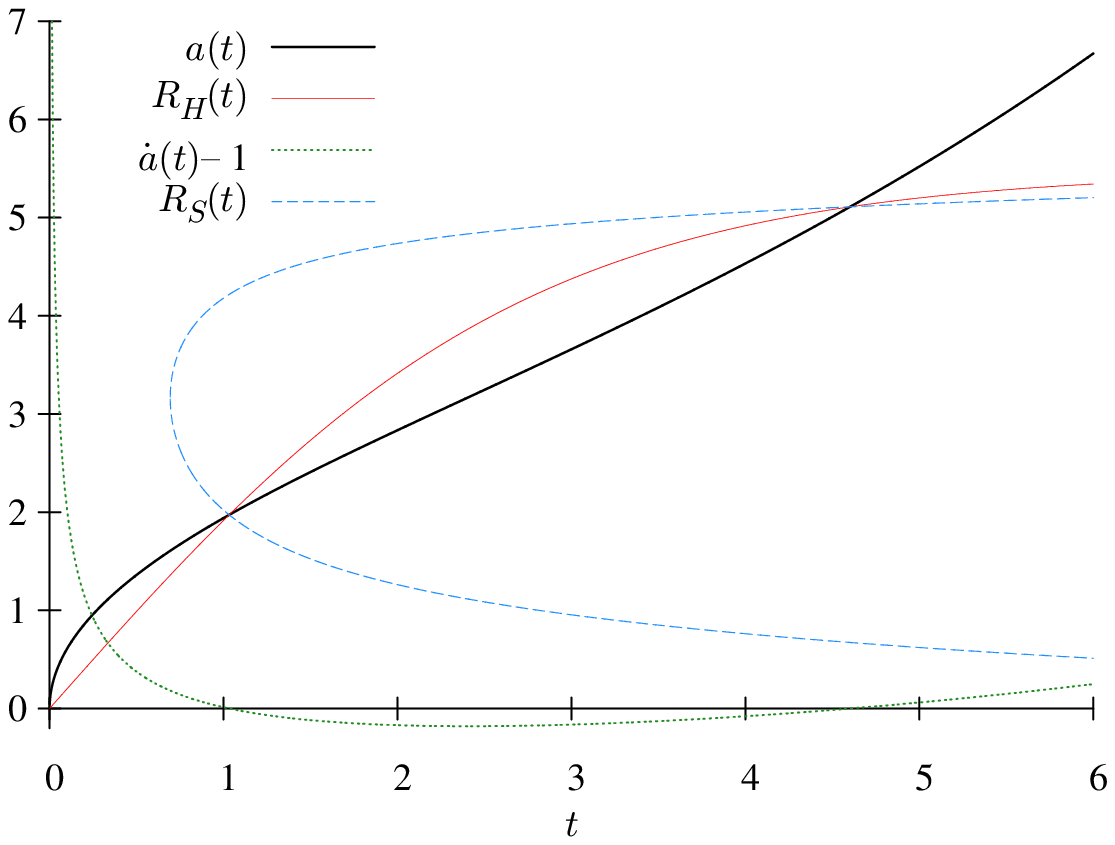}
\caption{Behaviour of quantities listed in Table~\ref{table:horizons}
  for the case $\Lambda \neq 0$. Top: dust ($w=0$), bottom: radiation
  ($w=\tfrac{1}{3}$); in both cases $M_0=4$ and
  $\Lambda=0.1$.\label{fig:horizons2}}
\end{figure}

The behaviour of the quantities listed in Table~\ref{table:horizons} for $\Lambda\neq 0$ are illustrated in Figure~\ref{fig:horizons2}.
In both cases, the values of $M_0$ and $\Lambda$ have been chosen so that, at least for some values of $t$, the condition $1-9M^2(t)\sqrt{\Lambda} > 0$ is satisfied and hence there exist two positive solutions for $R_{\rm S}(t)$, which correspond to the Schwarzschild radius and the de Sitter radius, respectively. 

In the case of dust (top panel), the mass-energy $M(t)$ enclosed within the spherical boundary $a(t)$ is constant, as expected, and thus so too are the two solutions for $R_{\rm S}(t)$. Remarkably, one again sees that each solution for $R_{\rm S}(t)$ intersects with $a(t)$ and $R_{\rm H}(t)$ at a single point, and that at both these intersections one has $\dot{a}(t)=1$. Thus, as in the case $\Lambda=0$, the comoving radius $a(t)$ initially lies outside the Hubble radius and enters it at precisely the same moment as it exits the Schwarzschild radius, at which point the fluid at the boundary $a(t)$ is moving at speed $c$. In the presence of non-zero $\Lambda$, however, the boundary $a(t)$ later exits the Hubble radius again, at precisely the same moment that it enters the de Sitter radius, at which point the fluid at the boundary is again moving with speed $c$. It is again worth noting in this dust case that the absence of pressure allows one consistently to `cut-off' the fluid at the boundary $a(t)$, and thereby consider a finite fluid ball surrounded by vacuum with $\Lambda \neq 0$, and obtain identical results.

The same generic behaviour to that outlined above is also seen for radiation in the bottom panel of Figure~\ref{fig:horizons2}. In this case, however, the non-zero pressure means that the fluid does work as it expands and so the mass-energy $M(t)$ contained within $a(t)$ decreases with time. Consequently, one initially has $1-9M^2(t)\sqrt{\Lambda} < 0$ and hence no positive solution for $R_{\rm S}(t)$. As $M(t)$ decreases, however, one eventually has $1-9M^2(t)\sqrt{\Lambda} > 0$ and so obtains two positive solutions for $R_{\rm S}(t)$, which again correspond to the Schwarzschild and de Sitter radii, respectively.

\section{Discussion and conclusions}
\label{sec:conc}

We have presented a comparison of our tetrad-based methodology for solving the Einstein field equations for spherically-symmetric systems with the traditional Lema\^{i}tre--Tolman--Bondi (LTB) model. Although the LTB model is widely used, it has a number of limitations. In particular, in its usual form it is restricted to pressureless systems. Moreover, the LTB model is typically expressed in comoving coordinates and thus provides a Lagrangian picture of the fluid evolution that can be difficult to interpret. Perhaps most importantly, however, even in the absence of vacuum regions the LTB metric contains a residual gauge freedom that necessitates the imposition of arbitrary initial conditions to determine the system evolution. As a consequence, we have for some time adopted a different, tetrad-based method for solving the Einstein field equations for spherically-symmetric systems.  The method was originally presented in \cite{Lasenby1998} in the language of geometric algebra, and was recently re-expressed in the more traditional tetrad notation in \cite{NLH1,NLH3}. The advantages of the tetrad-based approach are that it can straightforwardly accommodate pressure, has no gauge ambiguities (except in vacuum regions) and is expressed in terms of a `physical' (non-comoving) radial coordinate. As a result, in contrast to the LTB model, the method has a clear and intuitive physical interpretation. Indeed, the gauge choices employed result in equations that are essentially Newtonian in form.  

In comparing our tetrad-based methodology with the LTB model, we have focussed particularly on the issues of gauge ambiguity and the use of comoving versus `physical' coordinate systems. We have also clarified the correspondences, where they exist, between the two approaches.  As an illustration, we applied both methods to the classic examples of the Schwarzschild and Friedmann--Robertson--Walker (FRW) spacetimes.  In the former, we demonstrate that although the tetrad-based and LTB approaches both employ synchronous time coordinates and are based on trajectories of radially-infalling particles released from rest at infinity, the two methods lead to very different results corresponding to the use of Painlev\'e--Gullstrand and Lema\^{i}tre coordinates, respectively. For the FRW spacetime, we find that the LTB approach leads one to work directly in terms of the scale factor, whereas the tetrad-based method leads naturally to a description in terms of the Hubble parameter, which is a measurable quantity. Moreover, considerable gauge-fixing was required in the LTB model to obtain a definite solution, but this was unnecessary in the tetrad-based approach.

We have previously applied our tetrad-based method to modelling the evolution of a finite-size, compensated, spherically-symmetric object with continuous radial density and velocity profiles that is embedded in an expanding background universe, assuming zero pressure throughout \cite{Lasenby1999,Dabrowski1999}. We have also previously used the approach to obtain solutions describing a point mass residing in an expanding universe containing a cosmological fluid with pressure \cite{NLH1}, and later a finite spherical region of uniform interior density embedded in a background of uniform exterior density, where the pressure may be non-zero in both regions \cite{NLH3}. To illustrate further the use of our tetrad-based approach, we here extended the analysis in \cite{NLH3} to a generalised form of `Swiss cheese' model, which consists of an interior spherical region surrounded by a spherical shell of vacuum that is embedded in an exterior background universe. In general, we allow the fluid in the interior and exterior regions to support pressure, and we do not demand that the interior region be compensated. We find that much of the analysis in \cite{NLH3}, including the specification of boundary conditions, can be applied with little modification, but additional care is needed in determining the solution in the vacuum region, which requires some gauge-fixing, as might be expected.

We paid particular attention to the form of the solution in the vacuum region and verified the validity of Birkhoff's theorem, the usual interpretation of which has recently been brought in question \cite{Zhang2012}. We also showed that the theorem holds not only at the level of the metric, but also directly in terms of the tetrad components. We compared our findings with those in \cite{Zhang2012} and re-examined their model system of a static, thin spherical mass surrounding a central point mass positioned at the origin. In particular, we verified that the form of the line-element in the vacuum region interior to the shell depends on both the mass and location of the shell, although this is unsurprising given that the shell puts the interior vacuum region into a deeper potential well with respect to infinity than would be the case in its absence.

The above investigations allowed us to re-examine critically the original theoretical arguments set out in \cite{Melia2007,Melia2009,MeliaShevchuk2012} for the so-called $R_{\rm h} = ct$ cosmological model, which has recently received considerable attention. After pointing out a number of objections to the $R_{\rm h} = ct$ based on recent observational data, we consider in particular the central assumption underlying the original theoretical argument for the model, namely that the comoving Hubble distance should be constant. We demonstrate that this is {\em not} required, and so find no reliable theoretical basis for the $R_{\rm h}= ct$ model.

These considerations in turn elucidated the behaviour of a number of `horizons' during the general-relativistic evolution of homogeneous and isotropic cosmological models. In particular, we considered the evolution of an imaginary spherical boundary of radius $a(t)$ that is comoving with the fluid and centred on some arbitrary origin. For a selection of analytical spatially-flat cosmological models, we compared $a(t)$ to the Schwarzschild and Hubble radii. In the case of vanishing cosmological constant, we find the generic behaviour (both for dust and radiation models) that the comoving radius $a(t)$ initially lies outside the Hubble radius, but eventually enters it at precisely the same moment as it exits the Schwarzschild radius; it is allowed to do the latter, since the fluid at the boundary is moving at speed $c$ at this instant.  Thus, one has two `horizon crossings' taking place simultaneously and in opposite directions. In the case $\Lambda\neq 0$, one can obtain two positive solutions for $R_{\rm S}(t)$, which correspond to the Schwarzschild radius and the de Sitter radius, respectively. One again finds that the comoving radius $a(t)$ initially lies outside the Hubble radius and enters it at precisely the same moment as it exits the Schwarzschild radius, at which point the fluid at the boundary $a(t)$ is moving at speed $c$. In the presence of non-zero $\Lambda$, however, the boundary $a(t)$ later exits the Hubble radius again, at precisely the same moment that it enters the de Sitter radius, at which point the fluid at the boundary is again moving with speed $c$. This interesting behaviour is not usually pointed out in the literature.

\medskip
\begin{acknowledgments}
DYK is supported by a Samsung Scholarship.
\end{acknowledgments}

%\appendix

\bibliography{refs}% Produces the bibliography via BibTeX.

\end{document}